\documentclass[letterpaper,english,reprint, aps, prl,superscriptaddress]{revtex4-2}
\usepackage[T1]{fontenc}
\usepackage[latin9]{inputenc}
\usepackage{amsmath}
\usepackage{amssymb}
\usepackage{graphicx}
\usepackage{xcolor}
\makeatletter

\pdfpageheight\paperheight
\pdfpagewidth\paperwidth

\usepackage{braket}

\makeatother

\usepackage{babel}
\begin{document}
\title{Field-induced Kitaev multipolar liquid in spin-orbit coupled $d^2$ honeycomb Mott insulators}
\author{Ahmed Rayyan}
\affiliation{Department of Physics, University of Toronto, Toronto ON M5S 1A7,
Canada}
\author{Derek Churchill}
\affiliation{Department of Physics, University of Toronto, Toronto ON M5S 1A7,
Canada}
\author{Hae-Young Kee}
\email{hykee@physics.utoronto.ca}

\affiliation{Department of Physics, University of Toronto, Toronto ON M5S 1A7,
Canada}
\affiliation{CIFAR Program in Quantum Materials, Canadian Institute for Advanced
Research, Toronto ON M5G 1M1, Canada}
\date{\today}
\begin{abstract}
The Kitaev model, characterized by bond-dependent Ising spin interactions among spin-orbit entangled dipole
moments in Mott insulators, offered a new approach to quantum spin liquids. Motivated by another type of
bond-dependent interaction among quadrupole moments in $5d^2$ Mott insulators, we provide a microscopic route
to uncover the Kitaev multipolar liquid, featuring fractionalized excitations out of non-Kramers doublets carrying
multipole moments. The key ingredient is the magnetic field that allows for bond-anisotropic quadrupoleoctupole interactions via mixing with the excited triplet states. The conditions to realize signatures of this phase
in real materials are also discussed.
\end{abstract}
\maketitle
\emph{Introduction\textemdash }Recently, there have been many studies on
candidate materials of Kitaev spin liquids (KSLs) as they offer
a platform for topological quantum computation \citep{Kitaev_2003,kitaev2006}.
The Kitaev honeycomb model consists of bond-dependent Ising interactions
leading to the KSL with Majorana fermion and $\mathbb{Z}_{2}$ vortex
excitations. It was shown that bond-dependent (or ``compass'') interactions
appear naturally in Mott insulators with strong spin-orbit coupling
since the spin sector of the localized wavefunctions becomes sensitive
to the orbital spatial orientation due to spin-orbit entanglement
\citep{jk2009prl,cjk2010prl,doi:10.1146/annurev-conmatphys-020911-125138,rau2014prl,https://doi.org/10.48550/arxiv.1303.5922,Rau_2016,Winter_2017,Motome_2020,doi:10.7566/JPSJ.90.062001,TREBST20221}. \textcolor{black}{Since then there has been an intensive search for candidate Kitaev materials described by an effective model of spin-orbit entangled $J_{\text{eff}}=1/2$ Kramers doublets \citep{plumb14,hskim15,koitzsch16,sandilands16,zhou16,Banerjee_2016,kim16,winter2016,Janssen_2017,Winter_2017,Wang_2017,Laurell2020,jk2009prl,cjk2010prl,singh2012,rau2014trigonal,HwanChun2015,williams16,PhysRevB.97.014407,sano2018,liu2020,PhysRevB.102.224429}.} 

Bond-dependent interactions are not limited to Kramers doublets; in
the $5d^{2}$ double perovskites, the $J=2$ states are further split
into a non-Kramers doublet and an excited triplet via $t_{2g}$-$e_{g}$
mixing \citep{Paramekanti_2020,Maharaj_2020,Voleti_2020}. The non-Kramers
doublet hosts quadrupole and octupole moments while lacking a dipole
moment, and the microscopic theory of the multipolar interactions
exhibit octupole-octupole and bond-dependent quadrupole-quadrupole
interactions \citep{Khaliullin_2021,Voleti_2021,Churchill_2022}.
Remarkably, such interactions on the honeycomb lattice take the form
of the extended Kitaev model, which includes the bond-dependent off-diagonal
exchanges $\Gamma$ and $\Gamma'$ along with the conventional Heisenberg
interaction. Given their similarity, one may question if there is
a way to realize the exactly-solvable Kitaev model in multipolar honeycomb
systems.

In this Letter, we present a microscopic theory to uncover the Kitaev
model among multipolar moments where non-Kramers doublets are fractionalized
into Majorana fermions and $\mathbb{Z}_{2}$ vortices; we call this
phase the \emph{Kitaev multipolar liquid} in analogy with the
KSL. The key ingredient to realize the KML is the application of a \textcolor{black}{magnetic field} which leads to bond-dependent quadrupole-octupole interactions ordinarily forbidden by time-reversal symmetry.
Below we first derive the low-energy effective
multipolar model including the time-reversal symmetry breaking terms
and present its classical phase diagram. Noticing a special point
in the phase diagram which maps to the pure antiferro-Kitaev model,
we investigate the extent of the KML in the quantum phase diagram
using exact diagonalization (ED) on the $24$-site cluster. We summarize
our results and discuss the conditions to realize \textcolor{black}{signatures of} the KML in $5d^{2}$
honeycomb insulators. 
\begin{figure}
\includegraphics[scale=0.165]{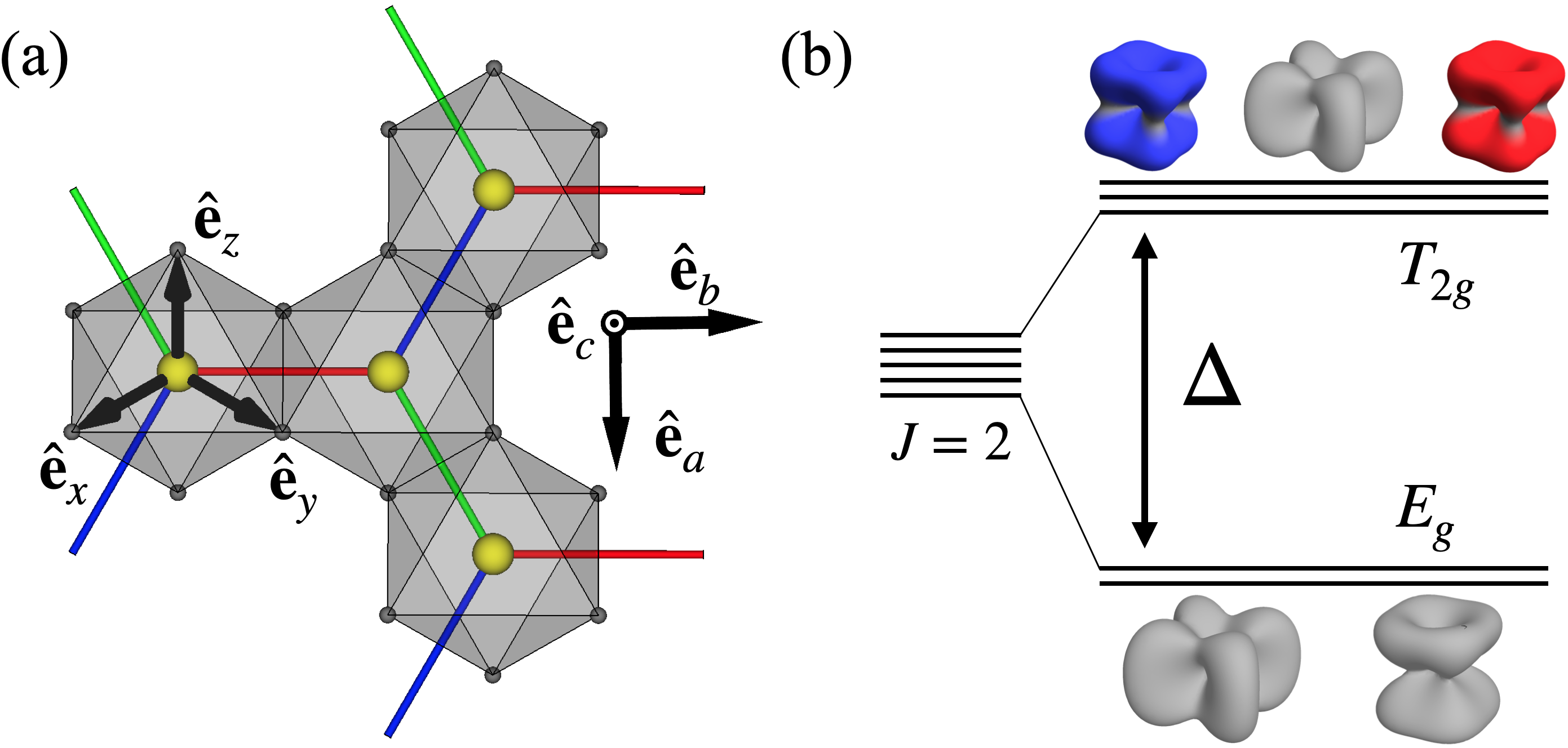}

\caption{\label{jeffandlattice} (a) The honeycomb lattice with transition
metal ions (shown in yellow) enclosed in an octahedral anion cage (shown
in grey). \textcolor{black}{The crystallographic $abc$ and octahedral $xyz$ coordinates are shown.} The $x,y,z$ bonds are colored green, blue, and red respectively.
(b) Single-ion level scheme for the $J=2$ moment. The fivefold degeneracy
is split by an energy gap $\Delta$ into a low-lying non-Kramers $E_{g}$
doublet and an excited $T_{2g}$ triplet by electronic $t_{2g}$-$e_{g}$
mixing induced by spin-orbit coupling \citep{Paramekanti_2020,Voleti_2020}.
The $E_{g}$ and $T_{2g}$ states are also shown where red and blue
represent non-zero spin density.}
\end{figure}

\emph{Multipolar pseudospin-1/2 interactions}\textemdash Electronic
states of transition metal ions enclosed in an octahedral cage are
generally split by cubic crystal fields into a low-lying $t_{2g}$
triplet and an excited $e_{g}$ doublet. For a $d^{2}$ filling, the
orbital sector is described by three antisymmetrized two-electron
states, forming an effective total angular momentum $L=1$ which is
then coupled to the total spin $S=1$ via spin-orbit coupling, resulting
in the $J=2$ multiplet \citep{Chen_2011}. The $J=2$ dipole operators
are given by $J_{\gamma}=\hat{\mathbf{e}}_{\gamma}\cdot\mathbf{J}$
for $\gamma\in\left\{ x,y,z\right\} $, where $\hat{\mathbf{e}}_{x,y,z}$
point along the three anion directions, see Fig. \ref{jeffandlattice}(a).
The fivefold-degenerate $J=2$ state can then be further split by
virtual processes mixing the electronic $t_{2g}$ and $e_{g}$ states
via spin-orbital excitations, resulting in a ground state doublet
and excited triplet separated by energy gap $\Delta$, see Fig. \ref{jeffandlattice}(b).
In analogy with the five electronic $d$ orbital states, we refer
to the doublet and triplet states as $E_{g}$ and $T_{2g}$ respectively.
The $E_{g}$ doublet is of the non-Kramers type with vanishing magnetic
dipole moment, but carries higher-rank moments i.e., quadrupole and
octupole moments denoted by the operators $Q_{x^{2}-y^{2}}=J_{x}^{2}-J_{y}^{2}$,
$Q_{3z^{2}}=\left(3J_{z}^{2}-\mathbf{J}^{2}\right)/\sqrt{3}$, and
$T_{xyz}=\frac{\sqrt{15}}{6}\overline{J_{x}J_{y}J_{z}},$ where the
overline symbol denotes symmetrization of the underlying operators.
Let us define three operators $s^{a,b,c}$ as
\begin{equation}
\left(s^{a},\,s^{b},\,s^{c}\right)\equiv\frac{1}{2}\mathcal{P}_{E_{g}}^{\dagger}\left(\frac{Q_{3z^{2}}}{2\sqrt{3}},\,\frac{Q_{x^{2}-y^{2}}}{2\sqrt{3}},\,\frac{T_{xyz}}{3\sqrt{5}}\right)\mathcal{P}_{E_{g}},\label{eq:pseudospindef}
\end{equation}
where $\mathcal{P}_{E_{g}}$ is the projection operator onto the $E_{g}$
doublet. The action of these operators on the $E_{g}$ subspace can
be represented by the three Pauli matrices $\left(s^{a},\,s^{b},\,s^{c}\right)=\left(\sigma^{3},\,\sigma^{1},\,\sigma^{2}\right)/2$,
so that $s^{a,b,c}$ form effective pseudospin-1/2 operators. \textcolor{black}{The
components are given by $s^{\bar{\gamma}}=\hat{\mathbf{e}}_{\bar{\gamma}}\cdot\mathbf{s}$
for $\bar{\gamma}\in\left\{ a,b,c\right\} $, where $\hat{\mathbf{e}}_{c}$
points out of the honeycomb plane spanned by $\hat{\mathbf{e}}_{a}$
and $\hat{\mathbf{e}}_{b}$, see Fig. \ref{jeffandlattice}(a).} The quadrupolar
and octupolar moments are in one-to-one correspondence with the projection
of $\mathbf{s}$ \textcolor{black}{onto the $ab$-plane or the $c$-axis,
respectively.}
\begin{figure}
\includegraphics[scale=0.12]{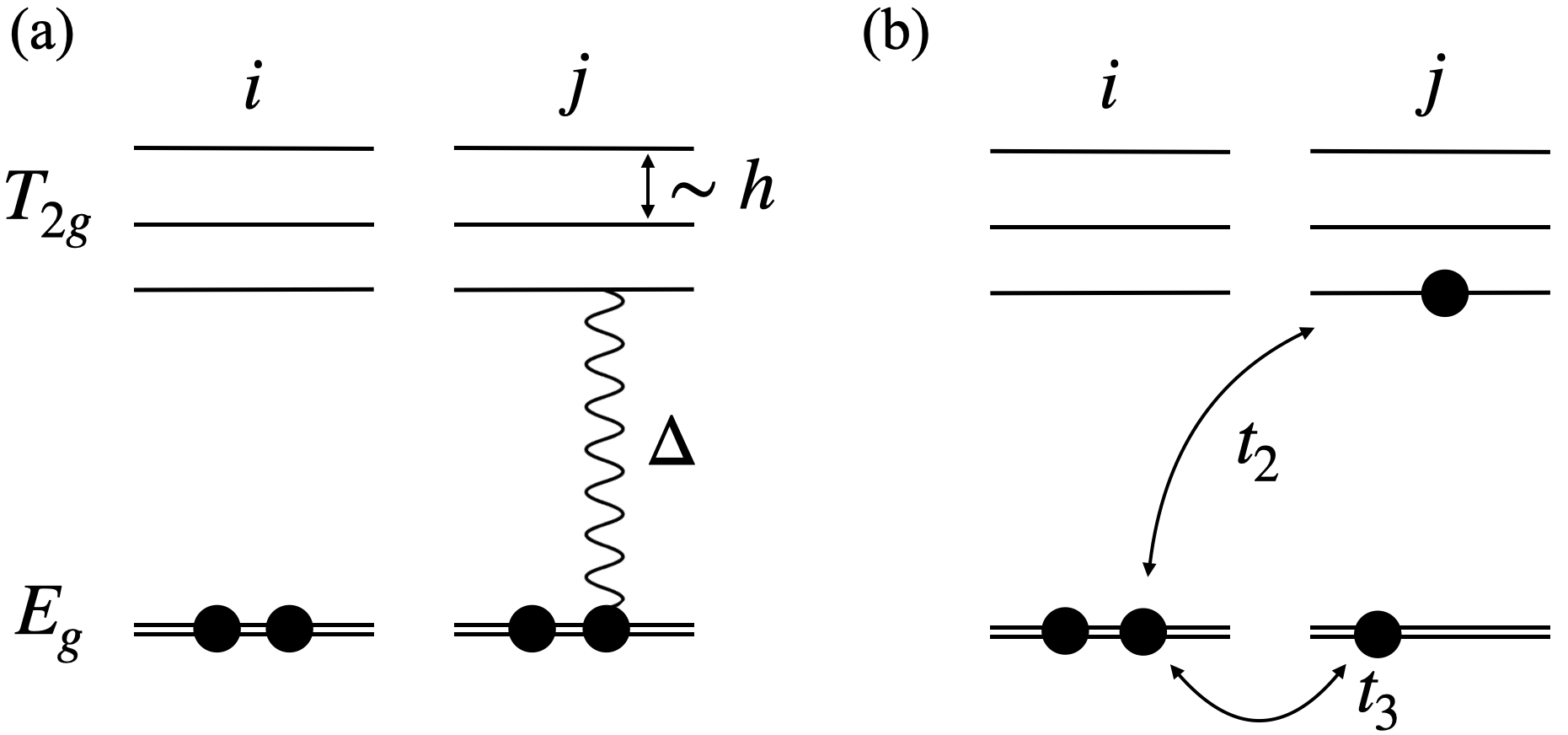}

\caption{\label{fig:hopdiagram} A \textcolor{black}{schematic of a} virtual process that contributes
to both $J_{B}$ and $h_{\text{eff}}$ term in the effective Hamiltonian
Eq. \eqref{eq:effH} at third order \textcolor{black}{for the case of a magnetic field aligned along the $c$-axis.} The $E_{g}$ and $T_{2g}$ states
at site $j$ mix due to (a) the on-site Zeeman field, and (b) hopping of an electron via interorbital
$t_{2}$ and intraorbital $t_{3}$, see visual representation of the
orbital overlaps in Fig. 1 of the SM \citep{SuppMat}. The overall contribution is
then proportional to $\left(t_{2}t_{3}/U\right)\left(h/\Delta\right)$.}
\end{figure}

We now investigate the form of the multipolar interactions by introducing
$t_{2g}$ orbital hopping, as was done in the case of the $d^{2}$
double perovskites \citep{Khaliullin_2021,Voleti_2021,Churchill_2022}.
On a honeycomb $z$-bond, the parameters $t_{3}$ and $t_{1}$ represent
intraorbital hopping through $xy-xy$ overlap, or $xz-xz$ and $yz-yz$
overlaps, respectively; see Fig. 1 of the Supplemental Material (SM) \citep{SuppMat}. We also introduce an $xz-yz$ interorbital
hopping through the edge-shared anions by hopping parameter $t_{2}$.
We go beyond earlier studies by immersing the system in an external
magnetic field \textcolor{black}{$\mathbf{h} = (h^x,\,h^y,\,h^z)$.} The spin and orbital degrees of freedom are
sensitive to this field via a Zeeman coupling $H^\text{Z}=\mu_{B}\left(\mathbf{L}+2\mathbf{S}\right)\cdot\mathbf{h}= \textcolor{black}{g_J\mu_B\mathbf{J}\cdot\mathbf{h}}$,
which introduces off-diagonal matrix elements between the doublet
and triplet states \citep{SuppMat}. We ensure that the
$E_{g}$ and $T_{2g}$ manifolds remain well-separated by considering
the low-field limit 
\textcolor{black}{$g_{J}\mu_{B}|\mathbf{h}|\ll\Delta$}
so that the perturbative
expansion is carried out in both 
\textcolor{black}{$|\mathbf{h}|/\Delta$}
and $t_{ij}^{2}/U$,
where $t_{ij}$ is some hopping between sites $i$ and $j$ and $U$
is the Hubbard energy cost of double-occupancy. \textcolor{black}{The external field gives rise
to new virtual processes where the $E_{g}$ doublet mixes with the
polarized $T_{2g}$ triplet during the hopping procedure, see Fig.
\ref{fig:hopdiagram}.} This process generates new terms in the effective
Hamiltonian denoted by \textcolor{black}{$J^\gamma_{B}$} and \textcolor{black}{$\mathbf{h}_{\text{eff}} = (h_\text{eff}^a,\,h_\text{eff}^b,\,h_\text{eff}^c)$} appearing at
third order in addition to the previously derived $J_{\tau},J_{Q}$,
and $J_{O}$:
\begin{align}
H & =\sum_{\langle ij\rangle_{\gamma}}J_{\tau}\,\tau_{i}^{\gamma}\,\tau_{j}^{\gamma}+J_{Q}\left(s_{i}^{a}\,s_{j}^{a}+s_{i}^{b}\,s_{j}^{b}\right)+J_{O}\,s_{i}^{c}\,s_{j}^{c}\nonumber \\
 & \qquad\textcolor{black}{-}\sqrt{2}\textcolor{black}{J^\gamma_{B}}\,\left(\tau_{i}^{\gamma}s_{j}^{c}+s_{i}^{c}\,\tau_{j}^{\gamma}\right)-\sum_{i}\textcolor{black}{\mathbf{h}_{\text{eff}}\cdot\mathbf{s}_i},\label{eq:effH}
\end{align}
where $\tau^{\gamma}\equiv s^{a}\,\text{cos }\phi_{\gamma}+s^{b}\,\text{sin }\phi_{\gamma}$
is a compass quadrupole operator with $\phi_{\gamma}=0,\,2\pi/3,\,4\pi/3$
for a given bond of type $\gamma=z,x,y$. Crucially, the addition
of a magnetic field supplements the Hamiltonian of Ref. \citep{Churchill_2022}
with terms ordinarily forbidden by time-reversal symmetry, including
a \textcolor{black}{bond-anisotropic} quadrupole-octupole interaction \textcolor{black}{$J^\gamma_{B}$}, \textcolor{black}{that is, $J_B^x$, $J_B^y$, and $J_B^z$ generally differ in strength along each bond. For the case of a $[111]$ magnetic field $\mathbf{h}=h\hat{\mathbf{e}}_c$, the $J_B^\gamma$ interaction becomes bond-isotropic with $J_B\equiv J_B^x = J_B^y = J_B^z$, and $h_\text{eff}^a = h_\text{eff}^b = 0, h_\text{eff} \equiv h_\text{eff}^c$, where
\begin{align}
J_{B} &= \frac{8}{9}\frac{t_{2}\left(2t_{1}+t_{3}\right)}{U}\frac{g_J \mu_B h}{\Delta} j_x^{\uparrow -},\nonumber \\
h_{\text{eff}} & =\frac{2}{3}\frac{t_{2}\left(t_{1}-t_{3}\right)}{U}\frac{g_J \mu_B h}{\Delta}j_z^{\uparrow \bar{0}}-24\frac{\left(g_J \mu_B h\right)^3}{\Delta^2}j_x^{\uparrow +}j_z^{++}j_x^{+\uparrow},\label{eq:fieldinducedparams}
\end{align}}
where $j_{\alpha}^{\mu\nu}\equiv \bra{\mu}J_{\alpha}\ket{\nu}$,
 $\ket{\uparrow}$
is one of the $E_{g}$ states and $\left\{ \ket{\pm},\ket{\bar{0}}\right\} $
are the three $T_{2g}$ states \citep{SuppMat}. \textcolor{black}{For the remainder of this Letter we focus on the case of a [111] magnetic field; the general form of $J_B^\gamma$ and $\mathbf{h}_{\text{eff}}$ for an arbitrary magnetic field direction are given in Section II of the SM \citep{SuppMat}, along with the expressions for $J_\tau, J_Q, \text{ and } J_O$ previously derived in Ref. \citep{Churchill_2022}.
}
\begin{figure}
\includegraphics[scale=0.125]{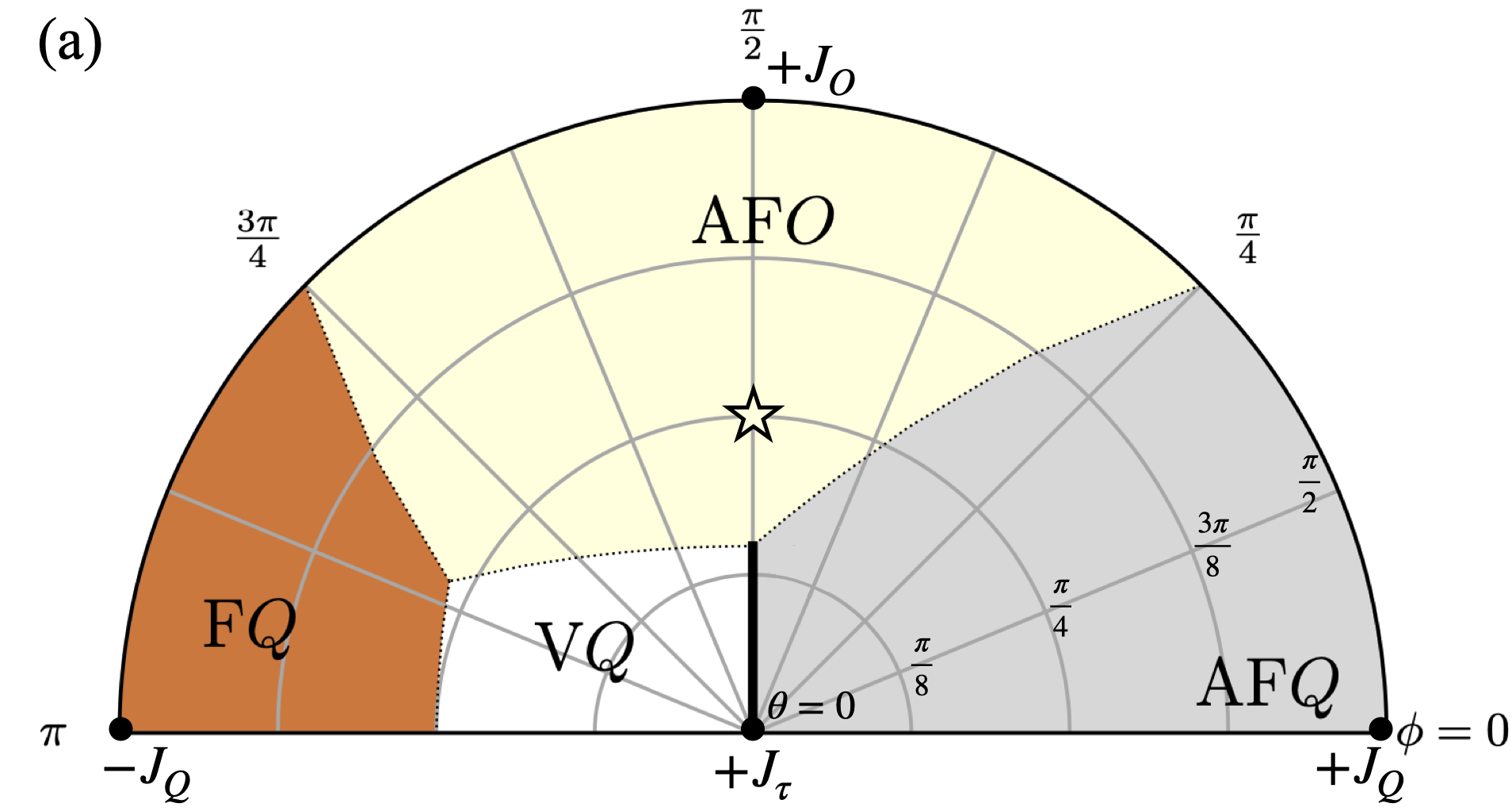}

\includegraphics[scale=0.165]{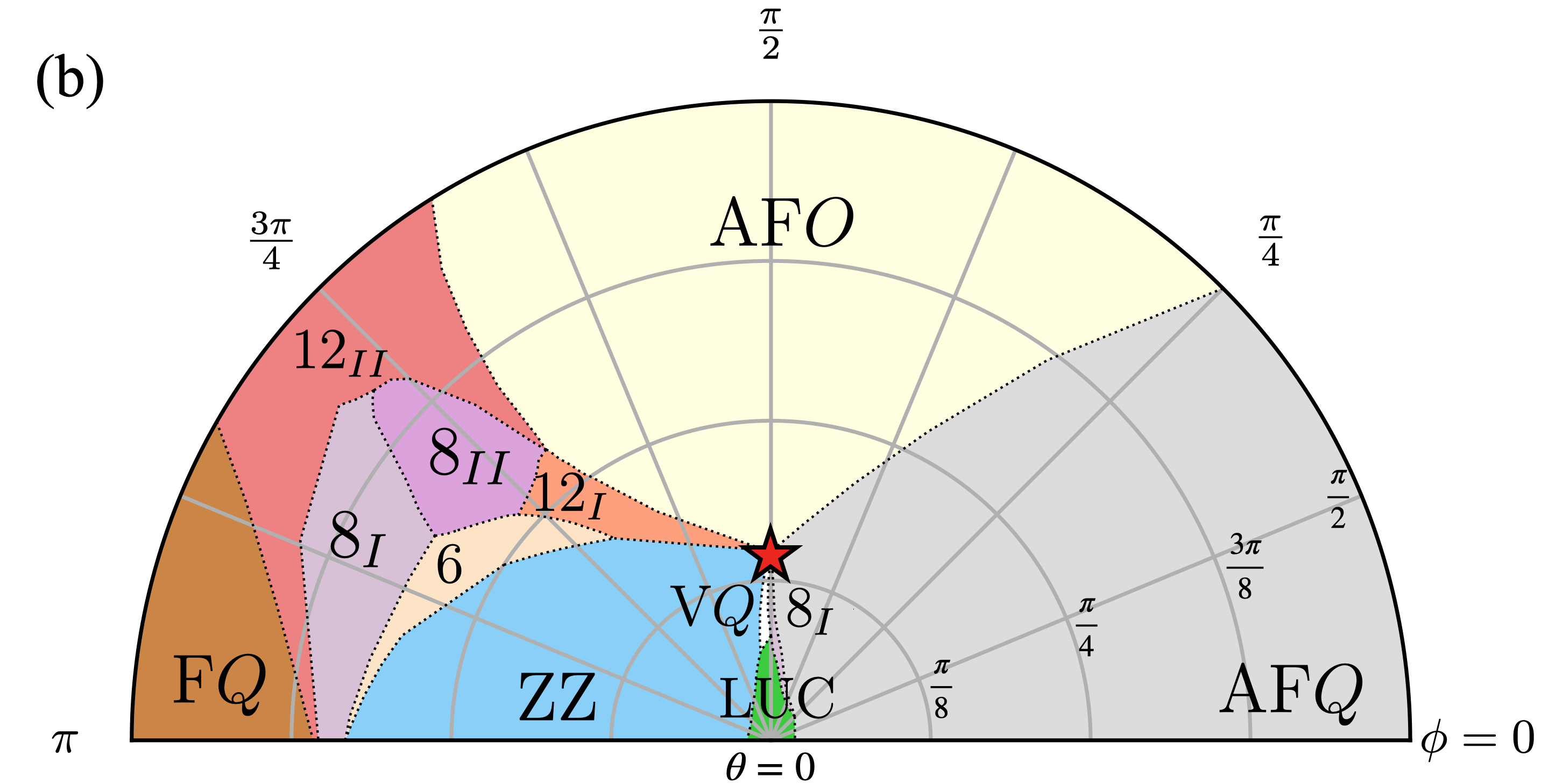}\caption{\label{fig:cpd} Classical phase diagrams computed by Monte Carlo
simulated annealing at $h_{\text{eff}}=0$ and (a) $J_{B}=0,$ and
(b) $J_{B}=\bar{J}/\sqrt{5}$, where $\bar{J}=1$ sets the energy
scale. The angles $\left(\theta,\,\phi\right)$ parameterize the exchange
interactions in Eq. \eqref{eq:effH} as $J_{\tau}=\bar{J}\,\text{cos }\theta,\,J_{Q}=\bar{J}\,\text{sin }\theta\text{ cos }\phi,\,J_{O}=\bar{J}\,\text{sin }\theta\text{ sin }\phi$. 
Some phases are labelled by the number of sites in the ordering unit
cell; \textcolor{black}{two such phases with identical unit cell size are distinguished using Roman numerals}. In Section III of the SM we
display the pseudospin configuration in each ordered phase \citep{SuppMat}. The yellow and red stars indicate points shown in Fig. 4; at the red star the Hamiltonian is equivalent to the pure antiferro-Kitaev model.}
\end{figure}

\emph{Classical phase diagrams\textemdash }We now explore the phase
diagram of the $J_{\tau}-J_{Q}-J_{O}-J_{B}-h_{\text{eff}}$ model.
We study the phase diagram of the Hamiltonian Eq. \eqref{eq:effH}
in the classical limit by treating $\mathbf{s}$ as an $O(3)$ vector
using Monte Carlo simulated annealing to obtain the classical ground
states \citep{metropolis53,kirkpatrick1983optimization,kirkpatrick1984optimization}; see Appendix A of Ref. \citep{Rayyan_2021} for simulation details. 
Signatures of quadrupolar and octupolar ordering are given by peaks
in the structure factors $\frac{1}{N}\sum_{ij}\left(s_{i}^{a}s_{j}^{a}+s_{i}^{b}s_{j}^{b}\right)\,e^{-i\mathbf{q}\cdot\left(\mathbf{r}_{i}-\mathbf{r}_{j}\right)}$
and $\frac{1}{N}\sum_{ij}s_{i}^{c}s_{j}^{c}\,e^{-i\mathbf{q}\cdot\left(\mathbf{r}_{i}-\mathbf{r}_{j}\right)}$,
respectively. We focus on the region where both $J_{\tau},J_{O}>0$
by setting $J_{\tau}=\bar{J}\,\text{cos }\theta,\,J_{Q}=\bar{J}\,\text{sin }\theta\text{ cos }\phi,\,J_{O}=\bar{J}\,\text{sin }\theta\text{ sin }\phi$
and restricting to $0\leq\theta\leq\pi/2$, $0\leq\phi\leq\pi$; for
$J_{B}=0$ the Hamiltonian is invariant under $\phi\rightarrow2\pi-\phi$
(i.e. $J_{O}\rightarrow-J_{O}$) and $s^{c}\rightarrow-s^{c}$ on
one of the two honeycomb sublattices. In Fig. \ref{fig:cpd}(a) we
present the phase diagram at fixed $J_{B}=h_{\text{eff}}=0$ which
is dominated by the antiferro-octupole (AF$O$), antiferro-quadrupole
(AF$Q$), ferro-quadrupole (F$Q$), and vortex-quadrupole (V$Q$)
phases; the V$Q$ phase in particular is a six-site quadrupolar phase,
see Section III of the SM for pseudospin configuration \citep{SuppMat}. Note that
each of these phases host either quadrupolar or octupolar moments,
but not both. The line where $J_{Q}=0$ and $0\leq J_{O}\leq J_{\tau}/2$
hosts a \emph{disordered }quadrupolar state originating from the pure
$J_{\tau}$ limit at $\theta=0$. There the model has a macroscopically
large ground state manifold owing to the physics of the $120^{\circ}$
compass honeycomb model \citep{Wu_2008,nasu2008,https://doi.org/10.48550/arxiv.1303.5922}.
The octupolar Ising interaction, which is proportional to $s_{i}^{c}s_{j}^{c}$,
does not immediately lift this degeneracy until $J_{O}>J_{\tau}/2$
where the AF$O$ phase is stabilized in a spin-flop transition. On
the other hand, the degeneracy is lifted by finite $J_{Q}$ and selects
either V$Q$ or AF$Q$ ordering depending on the sign of $J_{Q}.$ 

\begin{figure}
\includegraphics[scale=0.215]{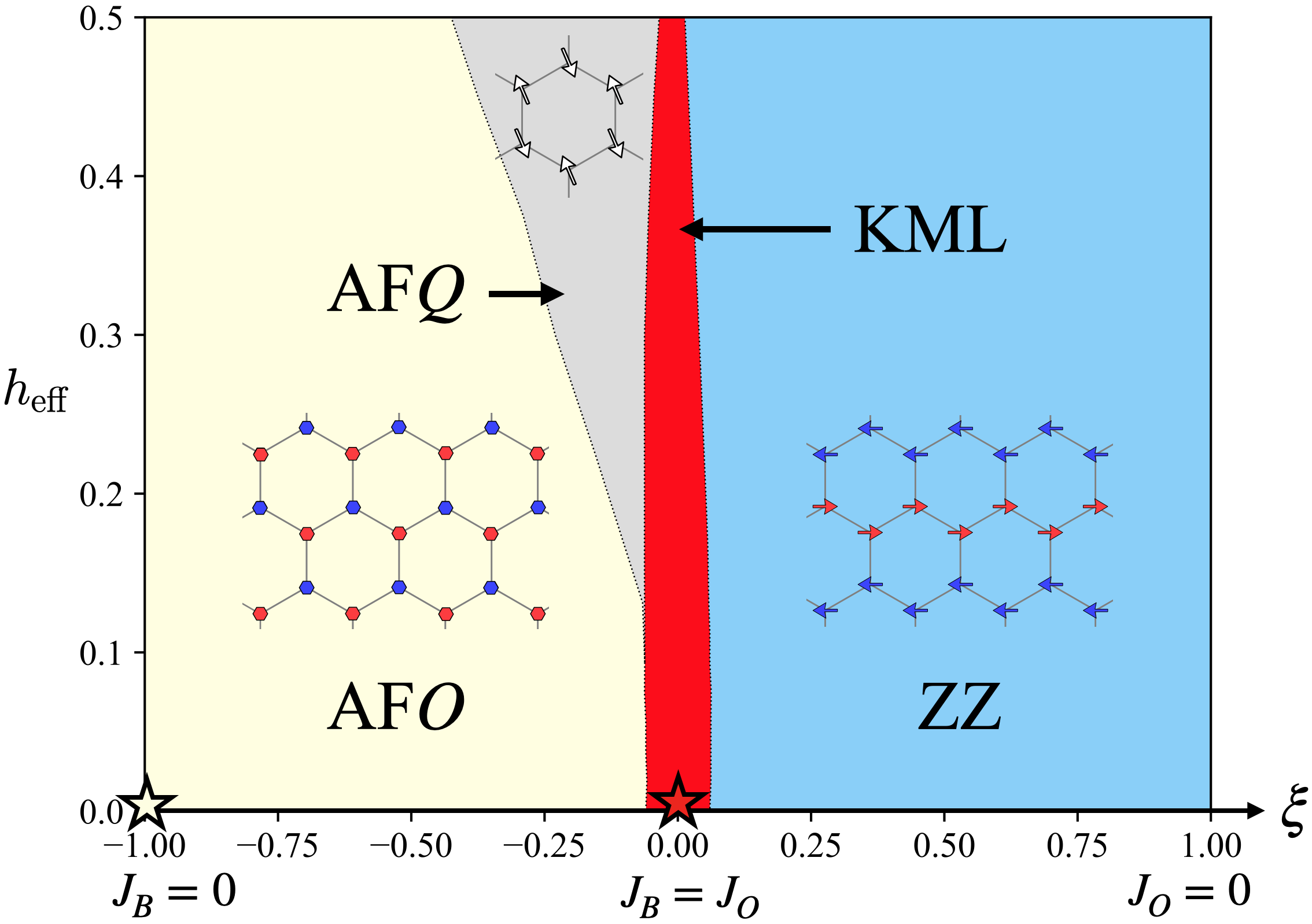}\caption{\label{fig:qpd} Quantum phase diagram obtained by 24-site ED, where
the parameters $\xi=\left(J_{B}-J_{O}\right)/\left(J_{B}+J_{O}\right)$
and $h_{\text{eff}}$ are tuned while $J_{Q}=0$ and $J_{\tau}=J_{B}+J_{O}=1$
are fixed. Phase boundaries are given by peaks in the ground state
energy derivatives, and we determine the presence and type of ordering
by calculating the quadrupolar and octupolar structure factors of
each phase shown in Section III of the SM \citep{SuppMat}. The yellow
star in corresponds to the point where $J_{\tau}=J_{O}$ whereas the
red star corresponds to the antiferro-Kitaev point. For each ordered
phase, the arrows represent each pseudospin's in-plane (i.e. quadrupolar)
component, whereas red and blue colors indicates the out-of-plane
(i.e. octupolar) component with opposite directions.}
\end{figure}

In Fig. \ref{fig:cpd}(b) we present the classical phase diagram at
a fixed value of $J_{B}=\bar{J}/\sqrt{5}>0$, which modifies the $J_{B}=0$
case in several notable ways. Firstly, the area surrounding the disordered\emph{
}quadrupolar state in the $J_{B}=0$ limit now hosts several large
unit cell (LUC) orders including 24-site and 40-site orders. Secondly,
whereas the region where both $J_{Q},J_{O}>0$ is relatively undisturbed,
the opposite limit where $J_{Q}$ and $J_{O}$ differ by a sign hosts
a variety of new ordered phases. An example is the zigzag (ZZ) phase
which contains both an in-plane and out-of-plane component, see Fig.
\ref{fig:qpd}. In fact, all new phases appearing in Fig. \ref{fig:cpd}(b)
feature both quadrupolar and octupolar moments, see Section III of
the SM for a visual representation of the classical pseudospin moments \citep{SuppMat}.
Thirdly, six different phases emerge from a single point indicated
by a red star in Fig. \ref{fig:cpd}(b). In the next section we explore
this point in detail and consider the consequences for the quantum
pseudospin model.

\emph{Kitaev multipolar liquid\textemdash }To find the relation to
the Kitaev model, we rewrite the Hamiltonian Eq. \eqref{eq:effH}
in the octahedral $xyz$ coordinates, where it may be written in the
$JK\Gamma\Gamma'$ form:
\begin{align}
H & =\sum_{\langle ij\rangle_{\gamma}}J\,\mathbf{s}_{i}\cdot\mathbf{s}_{j}+Ks_{i}^{\gamma}s_{j}^{\gamma}+\Gamma\left(s_{i}^{\alpha}s_{j}^{\beta}+s_{i}^{\beta}s_{j}^{\alpha}\right) \\
 & \hspace{1.5em}+\Gamma'\left[s_{i}^{\gamma}s_{j}^{\alpha}+s_{i}^{\alpha}s_{j}^{\gamma}+\left(\alpha\rightarrow\beta\right)\right]-h_\text{eff}\sum_{i}\hat{\mathbf{e}}_c\cdot\mathbf{s}_i\nonumber
 \label{eq:effHoctbasis}
\end{align}
where $s^{\gamma}=\hat{\mathbf{e}}_{\gamma}\cdot\mathbf{s}$, $\gamma\in\left\{ x,y,z\right\} $,
and $\alpha,\beta\in\left\{ x,y,z\right\} \backslash\left\{ \gamma\right\} $.
The values of $J,K,\Gamma$ and $\Gamma'$ are given by
\begin{equation}
\begin{aligned}
\label{eq:JKGGpparams}
J&= \, \frac{1}{3}\left(\frac{1}{2}J_{\tau}-2J_{B}+J_{O}+2J_{Q}\right), \\
K&=  \,\frac{1}{2}J_{\tau}+2J_{B},\quad \Gamma = J - J_Q, \\
\Gamma'&=  \frac{1}{3}\left(-J_{\tau}+J_{B}+J_{O}-J_{Q}\right).
\end{aligned}
\end{equation}
The special point indicated by the red star in Fig. \ref{fig:cpd}(b)
is where $J_{Q}=h_{\text{eff}}=0$ and the other parameters satisfy
the ratio $J_{\tau}:J_{O}:J_{B}=2:1:1$. Here the Hamiltonian takes
the form $H=\sum_{\langle ij\rangle_{\gamma}}\bar{K}\,s_{i}^{\gamma}s_{j}^{\gamma}$
where $\bar{K}=3J_{\tau}/2>0$; in other words, our multipolar pseudospin
model is described purely by an antiferro-Kitaev interaction. The
classical limit of this model hosts an extensive ground state degeneracy,
which explains why several classical phases meet at the red star in
Fig. \ref{fig:cpd}(b). In analogy to the Kitaev honeycomb model for
spin-1/2 moments, we can write the multipolar pseudospin operator
in terms of Majorana fermions $b^{\gamma}$ and $c$ as $s^{\gamma}=ib^{\gamma}c/2$,
and the model can be solved exactly in terms of Majorana fermions
hopping with a Dirac dispersion in the presence of a background $\mathbb{Z}_{2}$
gauge field. The resulting entangled ground state lacks long-range
multipolar order, which we recognize as the KML\emph{. }The discovery
of an exotic phase in an exactly-solvable model of multipolar moments
in $d^{2}$ honeycomb materials forms the central result of this Letter.

We would like to find the ordered multipole phases nearby the KML
phase space as $J_{B}$ and $h_{\text{eff}}$ are tuned, as the KML
physics may govern the finite temperature above which the multipole
ordering melts. To do so, we solve Eq. \eqref{eq:effH} using ED on
the $24$-site cluster with the numerical package $\mathcal{H}\Phi$
\citep{Kawamura_2017}. We parameterize the quantum phase diagram
by fixing $J_{Q}=0$ and tuning the parameter $\xi=\left(J_{B}-J_{O}\right)/\left(J_{B}+J_{O}\right)$
between $\xi\in[-1,1]$. Whereas $\xi=-1$ corresponds to the point
where $J_{B}=0$ and $J_{\tau}=J_{O}$, shown in Fig. \ref{fig:cpd}(a)
by the yellow star, $\xi=0$ corresponds to the point where $J_{B}=J_{O}=J_{\tau}/2$
which maps to the pure antiferro-Kitaev point. The three ordered phases
AF$O$, AF$Q$, and ZZ dominate the phase diagram. However, there
exists a narrow window where the KML is stabilized which is extended
in the $h_{\text{eff}}$ direction until roughly $h_{\text{eff}}\sim0.6\,J_{\tau}$.
At the antiferro-Kitaev point, the KML is not immediately susceptibile
to polarization as the ferro-octupole configuration does not lie within
the antiferro-Kitaev ground state manifold.

\textcolor{black}{One may expect that, due to the relations implied by Eq. \eqref{eq:JKGGpparams}, other ordered phases explored in the $JK\Gamma\Gamma'$ literature can be stabilized by the model Eq. \eqref{eq:effH} including the four-site stripy phase. Yet this phase does not appear in our work, as it primarily occupies the $K<0$ region \citep{cjk2010prl,rau2014prl} and we have focused on the region where $J_\tau,J_B\geq0$, ie. $K\geq0$. A small stripy phase could appear near $K>0$ when $\Gamma<0$ and $\Gamma'>0$ \citep{rau2014trigonal} but it lies outside our parameter space.} 

\emph{Discussion and summary\textemdash }We now discuss the conditions
to realize the KML in $5d^{2}$ insulators. The bond-dependent quadrupole-octupole
interaction requires the Zeeman field which induces off-diagonal components
between the $E_{g}$ and $T_{2g}$ states while maintaining their
energy separation $\Delta$; thus the first condition is that \textcolor{black}{$g_J\mu_B h\ll\Delta$. In the Os$^{6+}$ and Re$^{5+}$ double perovskites $\Delta$ is around
10-20 meV, restricting $h\sim\mathcal{O}\left(10\text{ T}\right)$. The second condition is approaching the Kitaev limit. In most edge-sharing materials, $|t_2|,|t_3|\gg |t_1|$ \citep{rau2014prl} so that the Kitaev limit is best approached if $t_2^2 \sim 2t_1 t_3$, since $J_Q\sim0$ and $J_O\sim 4t_2^2/3$ while $J_\tau\sim4t_3^2/9$, see expressions given in Section I of the SM \citep{SuppMat}. The ratio of $t_2$ and $t_3$ then determines whether the material is V$Q-$, AF$Q-$, or AF$O-$ordered in the zero-field limit shown in Fig. \ref{fig:cpd}(a). As the magnetic field is introduced and $J_B$ is increased, the system approaches the KML phase space but it may remain in the ordered phase depending on $t_2/t_3$. The signature of KML physics is then revealed at the finite temperatures above which the ordering vanishes.} The intricate balance between the field strength $h/\Delta$ and exchange paths presents a challenge for the material realization of the KML. Nevertheless, this work serves as a ``proof-of-concept'{}'' that the $d^{2}$ spin-orbit entangled honeycomb insulators with non-Kramers doublets may exhibit multipolar Kitaev physics.

\textcolor{black}{In this Letter} we have shown that the KML can arise in a spin-orbit coupled
$d^{2}$ honeycomb material. The key ingredient is the application
of a \textcolor{black}{magnetic field} which allows for bond-dependent
quadrupole-octupole interactions in the effective Hamiltonian of the
$E_{g}$ doublet. In combination with the Ising octupole and 120$^{\circ}$
compass-like quadrupole terms, the Hamiltonian can be tuned to the
pure Kitaev form. The resulting multipolar model can then be solved
exactly using Majorana fermions, in analogy with Kitaev\textquoteright s
original spin-1/2 model. Multipolar ordered phases arise along with
the KML, including those featuring combinations of quadrupolar and
octupolar ordering, as well as a disordered compass-quadrupole phase.
The nature and extent of these phases in the quantum phase diagram
form interesting avenues of future work. \textcolor{black}{We have also shown that the field-induced bond-dependent $J_B$ becomes bond-anisotropic when the field is tilted away from the $c$-axis. This extended phase space, and the novel physics contained within, motivates future studies of $d^2$ multipolar systems.}

\noindent \emph{Acknowledgements\textemdash }A.R. thanks P. P. Stavropoulos, S. Voleti, and F. D. Wandler for helpful discussions. We acknowledge support from the NSERC Discovery Grant No. 2022-04601. H.Y.K also acknowledges support from CIFAR and the Canada Research Chairs Program. Computations were performed on the Niagara supercomputer at the SciNet HPC Consortium. SciNet is funded by: the Canada Foundation for Innovation under the auspices of Compute Canada; the Government of Ontario; Ontario Research Fund - Research Excellence; and the University of Toronto.

\end{document}


\title{Supplemental material for ``Field-induced Kitaev multipolar liquid in spin-orbit coupled $d^2$ honeycomb Mott insulators''}
\author{Ahmed Rayyan}
\affiliation{Department of Physics, University of Toronto, Toronto ON M5S 1A7,
Canada}
\author{Derek Churchill}
\affiliation{Department of Physics, University of Toronto, Toronto ON M5S 1A7,
Canada}
\author{Hae-Young Kee}
\email{hykee@physics.utoronto.ca}

\affiliation{Department of Physics, University of Toronto, Toronto ON M5S 1A7,
Canada}
\affiliation{CIFAR Program in Quantum Materials, Canadian Institute for Advanced
Research, Toronto ON M5G 1M1, Canada}
\date{\today}
\maketitle
\section{\label{sec:tBEg}Derivation of the effective $E_{g}$ Hamiltonian}
In this section we outline how the effective $E_{g}$ pseudospin model
is obtained. We consider the limit where the octahedral crystal field
splitting and electron-electron repulsion are the dominant interactions,
justifying a strong-coupling approach where the $d^{2}$ electrons
lie in the $t_{2g}$ manifold and form a degenerate ground state that
is lifted by the electron hopping and magnetic field. Let us introduce
the creation operator $c_{m\sigma}^{\dagger}$ which creates a $t_{2g}$
electron with spin and orbital states denotes by $\sigma\in\left\{ +,-\right\} $
and $m\in\left\{ xy,xz,yz\right\} $, respectively. Our starting point
is the on-site Kanamori-Hubbard Hamiltonian on a single site $i$
which is given by
\begin{align}
H^{\text{KH}} & =U\sum_{m}n_{m+}n_{m-}+U'\sum_{m\neq m'}n_{m+}n_{m'-}+\left(U'-J_{H}\right)\sum_{m<m',\sigma}n_{m\sigma}n_{m'\sigma}\nonumber \\
 & +J_{H}\sum_{m\neq m'}\left(c_{m+}^{\dagger}c_{m'-}^{\dagger}c_{m-}c_{m'+}+c_{m+}^{\dagger}c_{m-}^{\dagger}c_{m'-}c_{m'+}\right)-\lambda\mathbf{L}\cdot\mathbf{S}\label{eq:kana}
\end{align}
where $n_{m\sigma}=c_{m\sigma}^{\dagger}c_{m\sigma}$ is the number
operator, and $\mathbf{L}=\boldsymbol{l}_{1}+\boldsymbol{l}_{2}$
and $\mathbf{S}=\boldsymbol{s}_{1}+\boldsymbol{s}_{2}$ are the total
orbital and spin angular momentum operators, respectively. The parameters
$U$ and $U'=U-2J_H$ are intraorbital and interorbital Coulomb interactions
respectively, where $J_H$ is the Hund's coupling. Lastly, the spin-orbit coupling parameter $\lambda$
is related to the single particle spin-orbit coupling $\zeta\left(\boldsymbol{l}_{1}\cdot\boldsymbol{s}_{1}+\boldsymbol{l}_{2}\cdot\boldsymbol{s}_{2}\right)$
via $\lambda=\zeta/n$ where $n=2$ is the electron filling. For the
$d^{2}$ filling, the limit $10Dq,U>\zeta,J_{H}$ leads to a fivefold
$J=2$ manifold that is further split into an $E_{g}$ doublet and
$T_{2g}$ triplet (see Section \ref{sec:J2} of Supplemental Material)
by spin-orbit-coupling induced $t_{2g}-e_{g}$ mixing with energy
gap $\Delta\sim\zeta^{2}/10Dq$ \citep{Paramekanti_2020,Voleti_2020}.

\begin{figure}[H]
\includegraphics[scale=0.245]{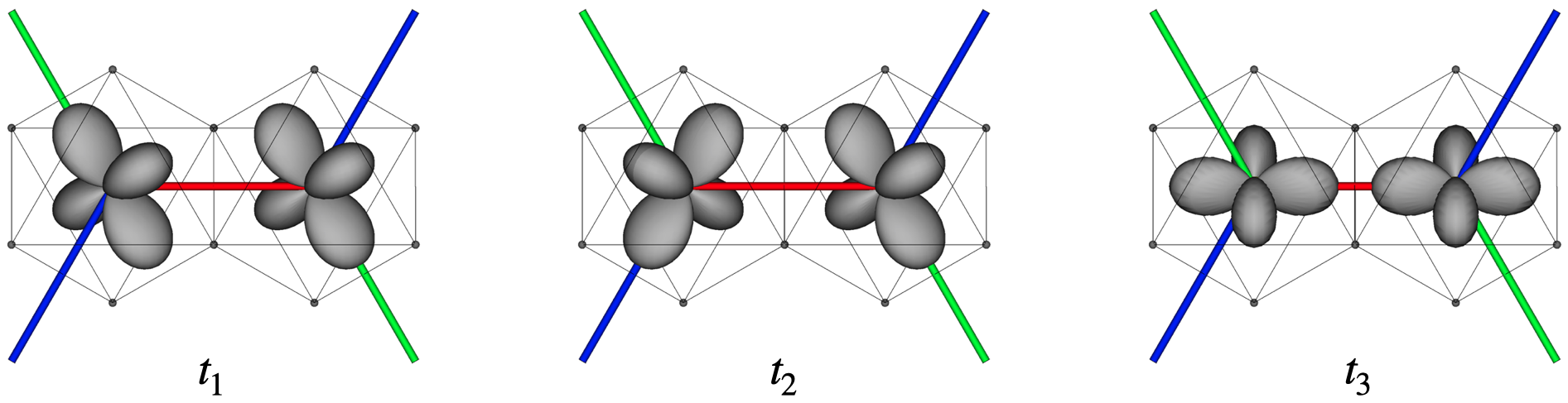}\caption{\label{fig:t2gorbitals}Examples of overlapping orbitals which contribute
to the hopping strengths $t_{1},t_{2},t_{3}$ in the $z$-bond hopping
matrix Eq. \eqref{eq:t2ghoppingmatrix}.}
\end{figure}

Now let us consider two honeycomb sites labelled $i=1,2$ connected
by a $z$-bond such that their octahedra share one edge. In the isolated
limit the Hamiltonian is given by $H_{1}^{\text{KH}}+H_{2}^{\text{KH}}$
with a ground state $E_{g}$ doublet on both sites. Interactions are
introduced by projecting the electron hopping and magnetic field onto
the $E_{g}$ doublet; in this section we focus on the electron hopping
as the case of the magnetic field is studied in the next section.
The orbital part of the tight-binding Hamiltonian is given by $H^{\text{TB}}_{\textcolor{black}{\langle12\rangle_{z}}} =c_{1}^{\dagger}T_{\langle12\rangle_{z}}c_{2}+c_{2}^{\dagger}T_{\langle12\rangle_{z}}^{\dagger}c_{1}$
where $c_{i}^{\dagger}=\left(c_{i,yz}^{\dagger},\,c_{i,xz}^{\dagger},\,c_{i,xy}^{\dagger}\right)$
and
\begin{equation}
T_{\langle12\rangle_{z}}=\left(\begin{array}{ccc}
t_{1} & t_{2} & 0\\
t_{2} & t_{1} & 0\\
0 & 0 & t_{3}
\end{array}\right)\label{eq:t2ghoppingmatrix}
\end{equation}
is fixed by inversion about the $z$-bond center and the $C_{2}$
symmetry about the $\hat{\mathbf{e}}_{b}$ axis for the undistorted
octahedra \citep{rau2014prl}. The hopping parameter $t_{3}$ and
$t_{1}$ correspond to direct intraorbital exchange, whereas $t_{2}$
parameterizes the effective interorbital $xz-yz$ hopping including
$p$-orbital superexchange. In Fig. \ref{fig:t2gorbitals} we display
the types of orbital overlap which lead to the three hopping parameters.

We now carry out the degenerate perturbation theory where the local
Kanamori-Hubbard Hamiltonian $H_{0}=H_{1}^{\text{KH}}+H_{2}^{\text{KH}}$
\textcolor{black}{
is perturbed by $V=\textcolor{black}{H^{\text{TB}}_{\textcolor{black}{\langle12\rangle_{z}}}+H^\text{Z}_{1}}+\textcolor{black}{H^\text{Z}_{2}}$, ie. the tight-binding Hamiltonian and the on-site Zeeman terms $H_i^\text{Z}=g_J\mu_B\mathbf{J}_i\cdot\mathbf{h}$ (see Section \ref{sec:J2} of Supplemental Material). The perturbative process generates quadratic and linear terms in $\mathbf{s}_i$ in the effective Hamiltonian. Along the $z$-bond, the quadratic terms take on the form
\begin{align}
H_{\langle12\rangle_{z}} & =J_{\tau}s_{1}^{a}s_{2}^{a}+J_{Q}\left(s_{1}^{a}s_{2}^{a}+s_{1}^{b}s_{2}^{b}\right)+J_{O}s_{1}^{c}s_{\textcolor{black}{2}}^{c}-\sqrt{2}\textcolor{black}{J^z_{B}}\left(s_{1}^{a}s_{2}^{c}+s_{1}^{c}s_{2}^{a}\right).\label{eq:effHzbond}
\end{align}}
\textcolor{black}{
where $s^{\bar{\gamma}}=\hat{\mathbf{e}}_{\bar{\gamma}}\cdot\mathbf{s}$
for $\bar{\gamma}\in\left\{ a,b,c\right\} $ are the pseudospin-1/2 components in the crystallographic $abc$ basis given by $\hat{\mathbf{e}}_{a}=\left(\hat{\mathbf{e}}_{x}+\hat{\mathbf{e}}_{y}-2\hat{\mathbf{e}}_{z}\right)/\sqrt{6}$, $\hat{\mathbf{e}}_{b}=\left(\hat{\mathbf{e}}_{y}-\hat{\mathbf{e}}_{x}\right)/\sqrt{2}$, and $\hat{\mathbf{e}}_{c}=\left(\hat{\mathbf{e}}_{x}+\hat{\mathbf{e}}_{y}+\hat{\mathbf{e}}_{z}\right)/\sqrt{3}$.}
The field-independent exchange parameters
are given by 
\begin{align}
J_{\tau} & =\frac{4}{9}\frac{(t_{1}-t_{3})^{2}}{U},\quad J_{Q}=\frac{2}{3}\frac{t_{1}(t_{1}+2t_{3})-t_{2}^{2}}{U},\quad J_{O}=\frac{2}{3}\frac{t_{1}(t_{1}+2t_{3})+t_{2}^{2}}{U},
\label{eq:fieldindepparams}
\end{align}
where we set $J_{H}$ to zero as it is small relative to $U$; the
derivation of $J_{\tau},J_{Q},\text{ and }J_{O}$ and the effect of
a finite Hund's coupling $J_{H}$ was previously studied in Ref. \citep{Churchill_2022}. The field-dependent \textcolor{black}{quadrupole-octupole interaction strength is given by 
\begin{equation}
\label{eq:fielddepparamsz}
\begin{aligned}
J_B^z&=\frac{8}{3\sqrt{6}}\frac{t_{2}\left(2t_{1}+t_{3}\right)}{U}\frac{g_{J}\mu_{B}h^z}{\Delta}.
\end{aligned}
\end{equation}
}
The quadratic terms along the $x$ and $y$ bonds can be obtained
via a $2\pi/3$ counter-clockwise rotation about the $[111]$ axis, and the pseudospin transforms accordingly as
\begin{equation}
\left(s^{a},s^{b},s^{c}\right)\rightarrow\left(-\frac{1}{2}s^{a}+\frac{\sqrt{3}}{2}s^{b},-\frac{\sqrt{3}}{2}s^{a}-\frac{1}{2}s^{b},s^{c}\right).
\end{equation}
To simplify notation it is convenient to introduce the compass quadrupole operators given by 
$
\tau^{\gamma}  \equiv s^{a}\,\text{cos }\phi_{\gamma}+s^{b}\,\text{sin }\phi_{\gamma}\text{ and }\bar{\tau}^{\gamma}  \equiv s^{b}\,\text{cos }\phi_{\gamma}-s^{a}\,\text{sin }\phi_{\gamma},
$
where $\phi_{\gamma}=0,\,2\pi/3,\,4\pi/3$ for a given bond of type
$\gamma=z,x,y$. The Hamiltonian on each honeycomb bond can then be
obtained by sending $s^{a}\rightarrow\tau^{\gamma}$ and $s^{b}\rightarrow\bar{\tau}^{\gamma}$
in Eq. \eqref{eq:effHzbond}. \textcolor{black}{Moreover, the quadrupole-octupole interaction is modified along each bond, namely
\begin{equation}\label{fielddepparamsxy}
J_B^x=\frac{8}{3\sqrt{6}}\frac{t_{2}\left(2t_{1}+t_{3}\right)}{U}\frac{g_{J}\mu_{B}h^x}{\Delta},\quad J_B^y=\frac{8}{3\sqrt{6}}\frac{t_{2}\left(2t_{1}+t_{3}\right)}{U}\frac{g_{J}\mu_{B}h^y}{\Delta}.
\end{equation}
The linear term in $\mathbf{s}_i$ is composed of bond-independent and bond-dependent terms; after summing over the contribution from each bond, the latter terms vanish.} The full effective Hamiltonian can then be written as 
\begin{equation}
H=\sum_{\langle ij\rangle_{\gamma}}J_{\tau}\tau_{i}^{\gamma}\tau_{j}^{\gamma}+J_{Q}\left(s_{i}^{a}s_{j}^{a}+s_{i}^{b}s_{j}^{b}\right)+J_{O}s_{i}^{c}s_{j}^{c}\textcolor{black}{-}\sqrt{2}\textcolor{black}{J^\gamma_{B}}\left(\tau_{i}^{\gamma}s_{j}^{c}+s_{i}^{c}\tau_{j}^{\gamma}\right)-\sum_{i}\textcolor{black}{\mathbf{h}_{\text{eff}}\cdot\mathbf{s}_i},
\label{eq:fulleffH}
\end{equation}
where we used the fact that $\tau_{i}^{\gamma}\tau_{j}^{\gamma}+\bar{\tau}_{i}^{\gamma}\bar{\tau}_{j}^{\gamma}=s_{i}^{a}s_{j}^{a}+s_{i}^{b}s_{j}^{b}$
for any bond $\gamma$, \textcolor{black}{ and $\mathbf{h}_\text{eff} = \left(h_\text{eff}^a,\,h_\text{eff}^b,\,h_\text{eff}^c\right)$ is expressed in the crystallographic basis $\hat{\mathbf{e}}_{a,b,c}$ where
\begin{equation}
\label{eq:effectivefielddepparams}
\begin{aligned}
h_\text{eff}^a&=6\left(g_J\mu_B\right)^2\frac{2h_z^2-h_x^2-h_y^2}{\Delta},\\
h_\text{eff}^b&=6\sqrt{3}\left(g_J\mu_B\right)^2\frac{h_x^2-h_y^2}{\Delta},\\
h_\text{eff}^c&=\frac{4}{3\sqrt{3}}\frac{t_{2}\left(t_{1}-t_{3}\right)}{U}\frac{g_{J}\mu_{B}\left(h^x+h^y+h^z\right)}{\Delta}-36\sqrt{3}\left(g_J\mu_B\right)^3\frac{h_x h_y h_z}{\Delta^2}.
\end{aligned}
\end{equation}
The main text is concerned with the case of a [111] magnetic field where $\mathbf{h}=h\,\hat{\mathbf{e}}_c = \frac{h}{\sqrt{3}}\left(\hat{\mathbf{e}}_x+\hat{\mathbf{e}}_y+\hat{\mathbf{e}}_z\right)$. In this case one finds that $h_\text{eff}^a = h_\text{eff}^b = 0$ and $J_B^x = J_B^y = J_B^z$, leading to a $C_3$-symmetric Hamtilonian. Defining $j_{\alpha}^{\mu\nu} = \bra{\mu}J_\alpha \ket{\nu}$ where $\mu,\nu$ label the $E_g$ and $T_{2g}$ states given in Section \ref{sec:J2} of the Supplemental Material, the field-dependent parameters are described by $J_B \equiv J_B^{x,y,z}$ and $h_\text{eff} \equiv h_\text{eff}^c$ given by
\begin{equation}
\begin{aligned}
J_B&=\frac{8}{9\sqrt{2}}\frac{t_{2}\left(2t_{1}+t_{3}\right)}{U}\frac{g_J \mu_B h}{\Delta},\\
&=\frac{8}{9}\frac{t_{2}\left(2t_{1}+t_{3}\right)}{U}\frac{g_J \mu_B h}{\Delta} j_x^{\uparrow -},\\
h_\text{eff}&=\frac{4}{3}\frac{t_{2}\left(t_{1}-t_{3}\right)}{U}\frac{g_{J}\mu_{B}h}{\Delta}-12\frac{\left(g_J \mu_B h\right)^3}{\Delta^2},\\
&=\frac{2}{3}\frac{t_{2}\left(t_{1}-t_{3}\right)}{U}\frac{g_J \mu_B h}{\Delta}j_z^{\uparrow \bar{0}}-24\frac{\left(g_J \mu_B h\right)^3}{\Delta^2}j_x^{\uparrow +}j_z^{++}j_x^{+\uparrow}.
\end{aligned}
\end{equation}}
\textcolor{black}{
The Hamiltonian Eq. \eqref{eq:fulleffH} can be written in a form which more clearly displays its symmetries by expanding the compass quadrupole operators $\tau^\gamma_i$, giving
\begin{align}
H & =\sum_{\langle ij\rangle_{\gamma}}\left(\frac{J_\tau}{2}+J_Q\right)\left(s_{i}^{a}s_{j}^{a}+s_{i}^{b}s_{j}^{b}\right)+J_{O}\,s_{i}^{c}s_{j}^{c}\\
 & \quad+\frac{J_\tau}{2}\left[\text{cos \ensuremath{\phi_{\gamma}}}\left(s_{i}^{a}s_{j}^{a}-s_{i}^{b}s_{j}^{b}\right)-\text{sin }\ensuremath{\phi_{\gamma}}\left(s_{i}^{a}s_{j}^{b}+s_{i}^{b}s_{j}^{a}\right)\right]\nonumber\\
 & \quad-\sqrt{2}J_B^\gamma\left[\text{cos \ensuremath{\phi_{\gamma}}}\left(s_{i}^{a}s_{j}^{c}+s_{i}^{c}s_{j}^{a}\right)+\text{sin }\ensuremath{\phi_{\gamma}}\left(s_{i}^{b}s_{j}^{c}+s_{i}^{c}s_{j}^{b}\right)\right]-\sum_i\mathbf{h}_\text{eff}\cdot\mathbf{s}_i\nonumber.
\end{align}
It is then straightforward to write the Hamiltonian in the octahedral basis $\hat{\mathbf{e}}_{x,y,z}$ since the pseudospin components $s^{\gamma}=\hat{\mathbf{e}}_{\gamma}\cdot\mathbf{s}$
for $\gamma\in\left\{ x,y,z\right\} $ are given by 
\begin{align*}
\left(\begin{array}{c}
s^{x}\\
s^{y}\\
s^{z}
\end{array}\right) & =\left(\begin{array}{ccc}
\frac{1}{\sqrt{6}} & -\frac{1}{\sqrt{2}} & \frac{1}{\sqrt{3}}\\
\frac{1}{\sqrt{6}} & \frac{1}{\sqrt{2}} & \frac{1}{\sqrt{3}}\\
-\frac{2}{\sqrt{6}} & 0 & \frac{1}{\sqrt{3}}
\end{array}\right)\left(\begin{array}{c}
s^{a}\\
s^{b}\\
s^{c}
\end{array}\right).
\end{align*}
By performing this rotation one obtains
\begin{equation}
\begin{aligned}
H & =\sum_{\langle ij\rangle_{\gamma}}J^\gamma\,\mathbf{s}_{i}\cdot\mathbf{s}_{j}+K^\gamma s_{i}^{\gamma}s_{j}^{\gamma}+\Gamma^\gamma \left(s_{i}^{\alpha}s_{j}^{\beta}+s_{i}^{\beta}s_{j}^{\alpha}\right)\\
&\qquad+\Gamma'^\gamma\left[s_{i}^{\gamma}s_{j}^{\alpha}+s_{i}^{\alpha}s_{j}^{\gamma}+\left(\alpha\rightarrow\beta\right)\right]-\sum_i\mathbf{h}_\text{eff}\cdot\mathbf{s}_i,
\end{aligned}
\end{equation}
where
\begin{align}
J^\gamma= & \frac{1}{3}\left(\frac{1}{2}J_{\tau}-2J_{B}^\gamma+J_{O}+2J_{Q}\right),\nonumber \\
K^\gamma= & \frac{1}{2}J_{\tau}+2J_{B}^\gamma,\quad \Gamma^\gamma = J^\gamma - J_Q,\nonumber \\
\Gamma'^\gamma= & \frac{1}{3}\left(-J_{\tau}+J_{B}^\gamma+J_{O}-J_{Q}\right).
\end{align}
For an external field along the $c$-axis, the exchange parameters are bond-isotropic, ie. $K^x = K^y = K^z$ and so on for $J,\,\Gamma,\,\text{and } \Gamma'$.}
\section{\label{sec:J2}Properties of $J=2$ states under a Zeeman field}
Here we discuss some details regarding the $J=2$ manifold, with angular
momentum states satisfying $J_{z}\ket{m_{J}}=m_{J}\ket{m_{J}}$ for
$m_{J}=-2,-1,0,1,2$. The $J=2$ splitting can be modelled by a residual
crystal field splitting $H_{\Delta}=\Delta\left(\mathcal{O}_{40}+5\mathcal{O}_{44}\right)/120$
where $\Delta>0$ and
\begin{equation}
\begin{aligned}
\mathcal{O}_{40} & =35J_{z}^{4}-\left(30J(J+1)-25\right)J_{z}^{2}+\text{const.}\\
\mathcal{O}_{44} & =(J_{+}^{4}+J_{-}^{4})/2,
\end{aligned}
\end{equation}
are Stevens operators \citep{Paramekanti_2020,Voleti_2020}. The eigenstates
of $H_{\Delta}$ form a doublet with energy $E=0$ and a triplet with
$E=\Delta>0$; these are the $E_{g}$ and $T_{2g}$ states, respectively.
We may choose the following basis for the $E_{g}$ states 
\begin{align}
\ket{\uparrow} & =\frac{1}{\sqrt{2}}\left(\ket{2}+\ket{-2}\right),\quad\ket{\downarrow}=\ket{0},
\end{align}
and the $T_{2g}$ states
\begin{align}
\ket{\pm} & =\ket{\pm1},\quad\ket{\bar{0}}=\frac{1}{\sqrt{2}}\left(\ket{2}-\ket{-2}\right).
\end{align}
Let us denote the projection onto these subspaces as $\mathcal{P}_{E_{g}}$
and $\mathcal{P}_{T_{2g}}$, respectively. The time-reversal operator
acts on the angular momentum states as $\mathcal{T}\ket{m_{J}}=\left(-1\right)^{m_{J}}\ket{-m_{J}}$
and on the angular momentum operator as $\mathcal{T}\mathbf{J}\mathcal{T}^{-1}=-\mathbf{J}$.
For either $E_{g}$ state $\ket{\sigma}$ where $\sigma\in\left\{ \uparrow,\downarrow\right\} ,$
time-reversal is equivalent to the identity $\mathcal{T}\ket{\sigma}=\ket{\sigma}$.
This implies that 
\begin{equation}
\begin{aligned}
\bra{\sigma}\mathbf{J}\ket{\sigma} & =\bra{\sigma}\mathcal{T}^{-1}\mathcal{T}\mathbf{J}\mathcal{T}^{-1}\mathcal{T}\ket{\sigma}\\
 & =-\bra{\sigma}\mathbf{J}\ket{\sigma},
\end{aligned}
\end{equation}
ie. $\bra{\sigma}\mathbf{J}\ket{\sigma}=0$: the $E_{g}$ doublet
does not carry a dipole moment. In fact it is carried by the $T_{2g}$
states; the three operators $l_{\gamma}=-\mathcal{P}_{T_{2g}}^{\dagger}J_{\gamma}\mathcal{P}_{T_{2g}}$
with $\gamma\in\left\{ x,y,z\right\} $ form the $l=1$ representation
of the $\mathfrak{su}\left(2\right)$ algebra. 

Now let us add a Zeeman field $\textcolor{black}{H^\text{Z}=\mu_{B}\left(\mathbf{L}+2\mathbf{S}\right)\cdot\mathbf{h}=g_{J}\mu_{B}\,\mathbf{J}\cdot\mathbf{h}}$,
where $g_{J}=1/2$ for $J=2$ by the Wigner-Eckart theorem. \textcolor{black}{In the $\left\{ \ket{\uparrow},\ket{\downarrow},\ket{-},\ket{+},\ket{\bar{0}}\right\} $
basis, $H^{\text{Z}}$ takes the form
\begin{align}
H^\text{Z} & =g_{J}\mu_{B}\left(\begin{array}{cc|ccc}
0 & 0 & \frac{1}{\sqrt{2}}h^+ & \frac{1}{\sqrt{2}}h^- & 2h_z\\
0 & 0 & \frac{\sqrt{3}}{2}h^- & \frac{\sqrt{3}}{2}h^+ & 0\\
\hline \frac{1}{\sqrt{2}}h^- & \frac{\sqrt{3}}{2}h^+ & -h_z & 0 & -\frac{1}{\sqrt{2}}h^-\\
\frac{1}{\sqrt{2}}h^+ & \frac{\sqrt{3}}{2}h^- & 0 & h_z & \frac{1}{\sqrt{2}}h^+\\
2h_z & 0 & -\frac{1}{\sqrt{2}}h^+ & \frac{1}{\sqrt{2}}h^- & 0
\end{array}\right).
\end{align}
where $\mathbf{h}=(h_x,h_y,h_z)$ is expressed in the octahedral basis $\hat{\mathbf{e}}_{x,y,z}$, and $h^{\pm} = h_x \pm i h_y$.}
Note that the Zeeman field leads to a finite off-diagonal block
connecting the $E_{g}$ and $T_{2g}$ manifolds. To understand how
the $E_{g}$ doublet are perturbed by \textcolor{black}{a [111]} magnetic field \textcolor{black}{$\mathbf{h}=h\,\hat{\mathbf{e}}_c$} we solve
the Hamiltonian $H_{\Delta}+H_{\text{Z}}$ in the limit where $\epsilon\equiv g_{J}\mu_{B}h/\Delta\ll1$
using the method of resolvents \citep{messiah75}; the modified $E_{g}$
states are given by (neglecting normalization constants)
\begin{equation}
\begin{aligned}
\ket{\tilde{\uparrow}}= & \ket{\uparrow}\textcolor{black}{-}\frac{\epsilon}{\sqrt{3}}\left(e^{-i\pi/4}\ket{-}+e^{i\pi/4}\ket{+}+2\ket{\bar{0}}\right)+\mathcal{O}\left(\epsilon^{2}\right)\\
\ket{\tilde{\downarrow}}= & \ket{\downarrow}\textcolor{black}{-}\epsilon\left(e^{i\pi/4}\ket{-}+e^{-i\pi/4}\ket{+}\right)+\mathcal{O}\left(\epsilon^{2}\right)
\end{aligned}
\end{equation}
with the field lowering the energy of both states to $E=-2\epsilon^{2}+\mathcal{\mathcal{O}}\left(\epsilon^{3}\right)$.

\section{Properties of multipolar orders}
In this section we discuss some properties of the multipolar phases
found in the phase diagrams of the main text.
\subsection{Phases in the classical model}
\begin{figure}[H]
\centering{}\includegraphics[scale=0.18]{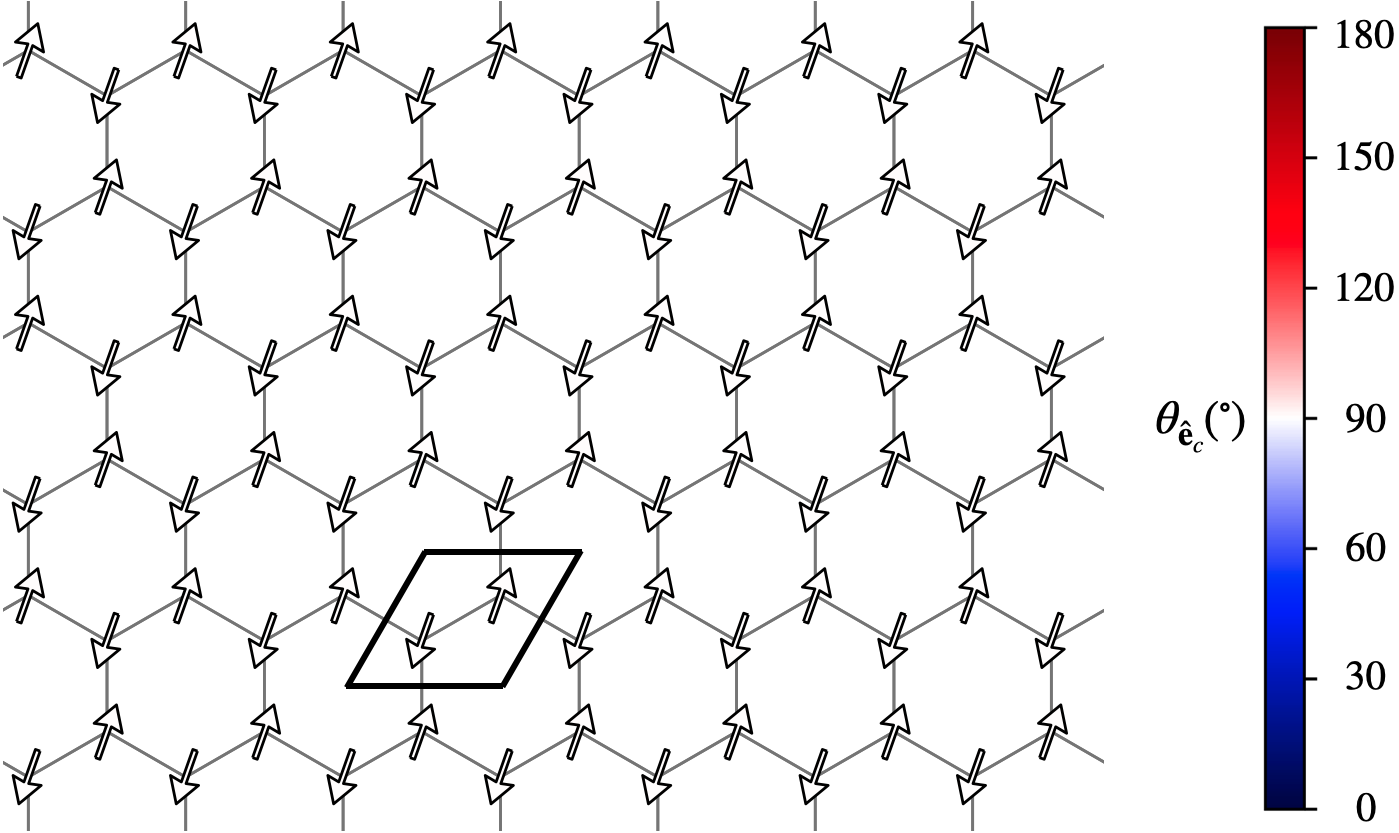}\caption{\label{fig:afqclassical} Pseudospin configuration at $\left(\theta/\pi,\,\phi/\pi,\,J_{B}/\bar{J}\right)=$
$\left(0.25,\,0.25,\,0.0\right)$ located within the AF$Q$ phase.
The colorbar, which indicates the size of the octupolar moment, applies
to Figs. \ref{fig:afqclassical}-\ref{fig:40LUCclassical}.}
\end{figure}

\begin{figure}[H]
\centering{}\includegraphics[scale=0.18]{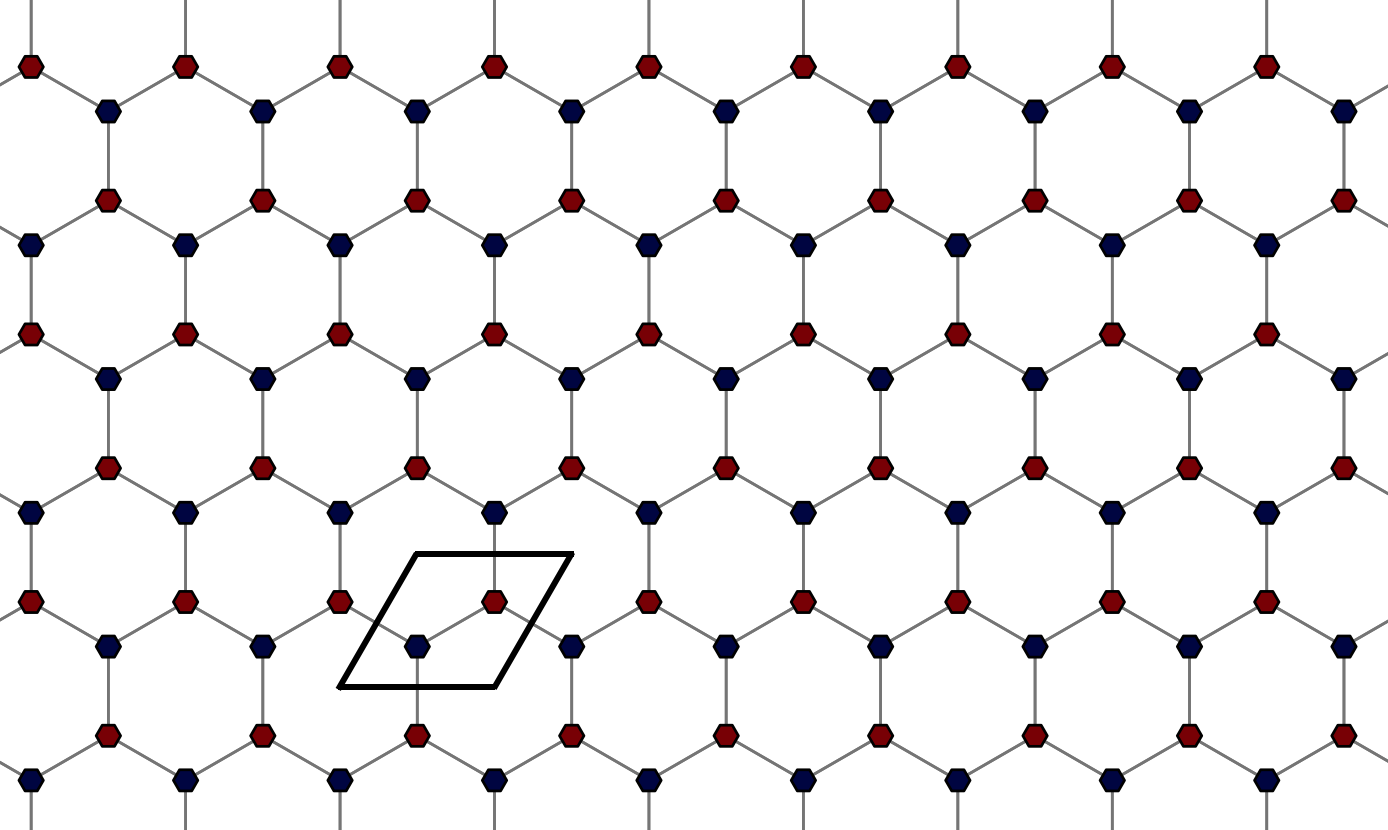}\caption{\label{fig:afoclassical} Pseudospin configuration at $\left(\theta/\pi,\,\phi/\pi,\,J_{B}/\bar{J}\right)=$
$\left(0.25,\,0.5,\,0.0\right)$ located within the AF$O$ phase.}
\end{figure}

\begin{figure}[H]
\centering{}\includegraphics[scale=0.18]{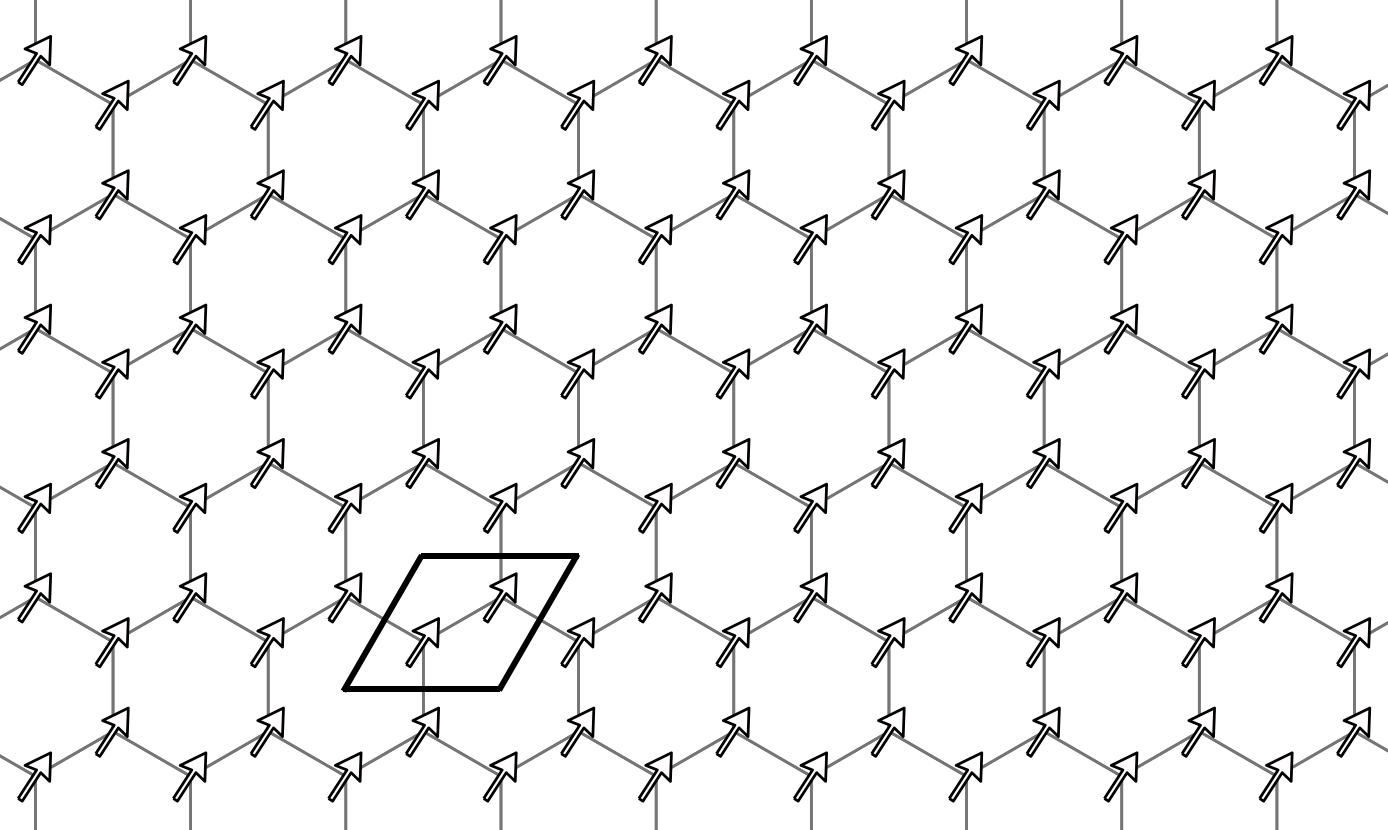}\caption{\label{fig:fqclassical} Pseudospin configuration at $\left(\theta/\pi,\,\phi/\pi,\,J_{B}/\bar{J}\right)=$
$\left(0.375,\,1.0,\,0.0\right)$ located within the F$Q$ phase.}
\end{figure}

\begin{figure}[H]
\centering{}\includegraphics[scale=0.18]{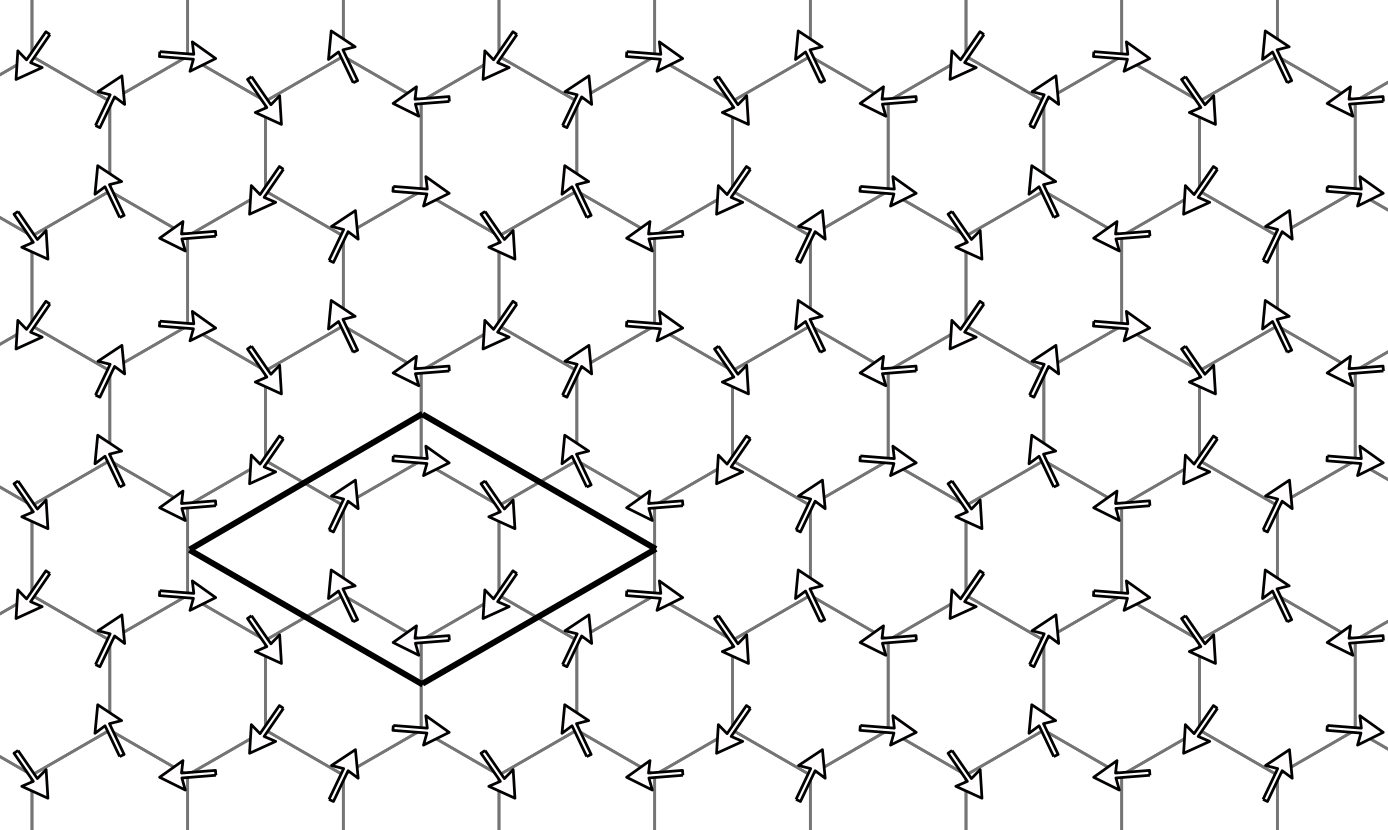}\caption{\label{fig:vqclassical} Pseudospin configuration at $\left(\theta/\pi,\,\phi/\pi,\,J_{B}/\bar{J}\right)=$
$\left(0.125,\,1.0,\,0.0\right)$ located within the V$Q$ phase.}
\end{figure}

\begin{figure}[H]
\centering{}\includegraphics[scale=0.18]{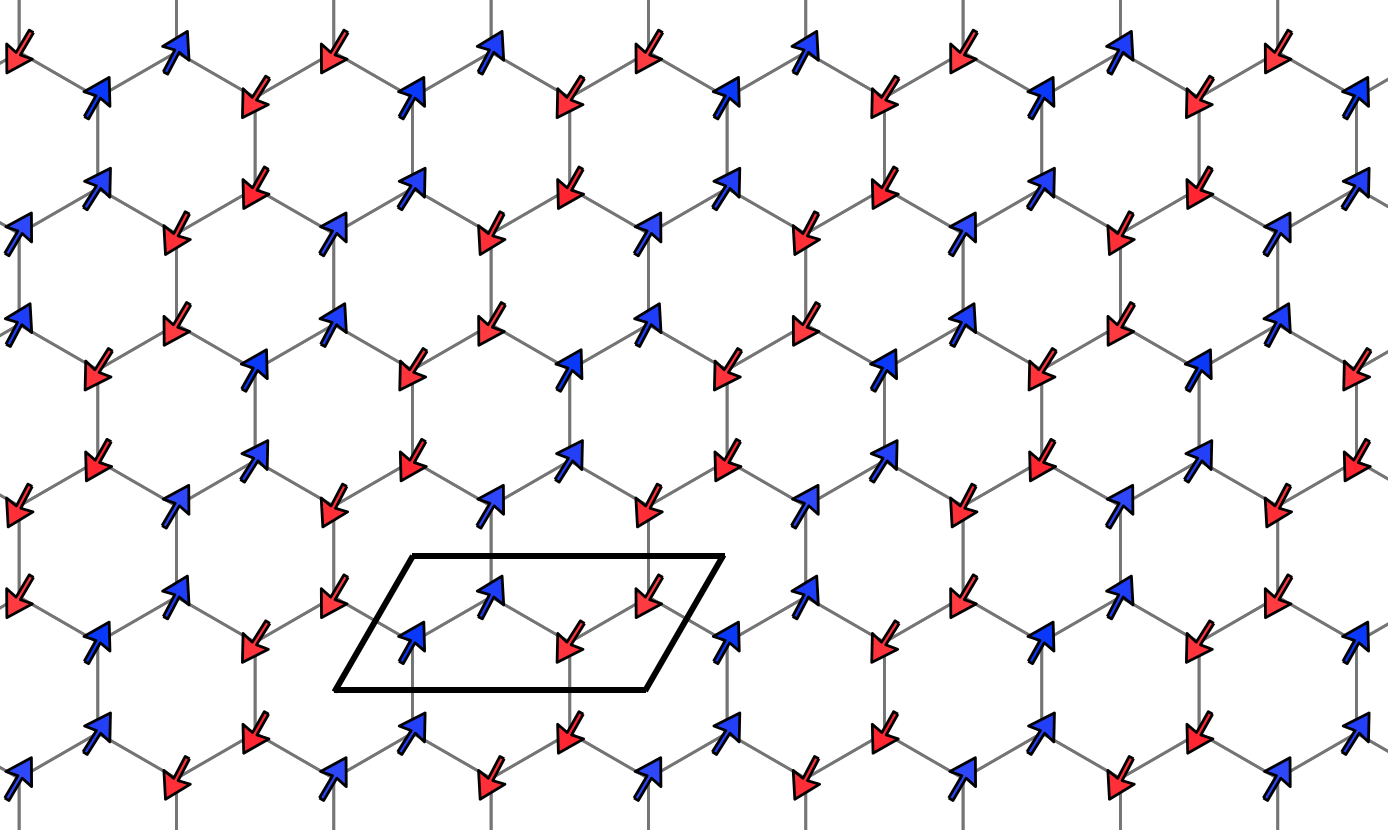}\caption{\label{fig:zzclassical} Pseudospin configuration at $\left(\theta/\pi,\,\phi/\pi,\,J_{B}/\bar{J}\right)=$
$\left(0.25,\,1.0,\,1/\sqrt{5}\right)$ located within the ZZ phase.}
\end{figure}

\begin{figure}[H]
\centering{}\includegraphics[scale=0.18]{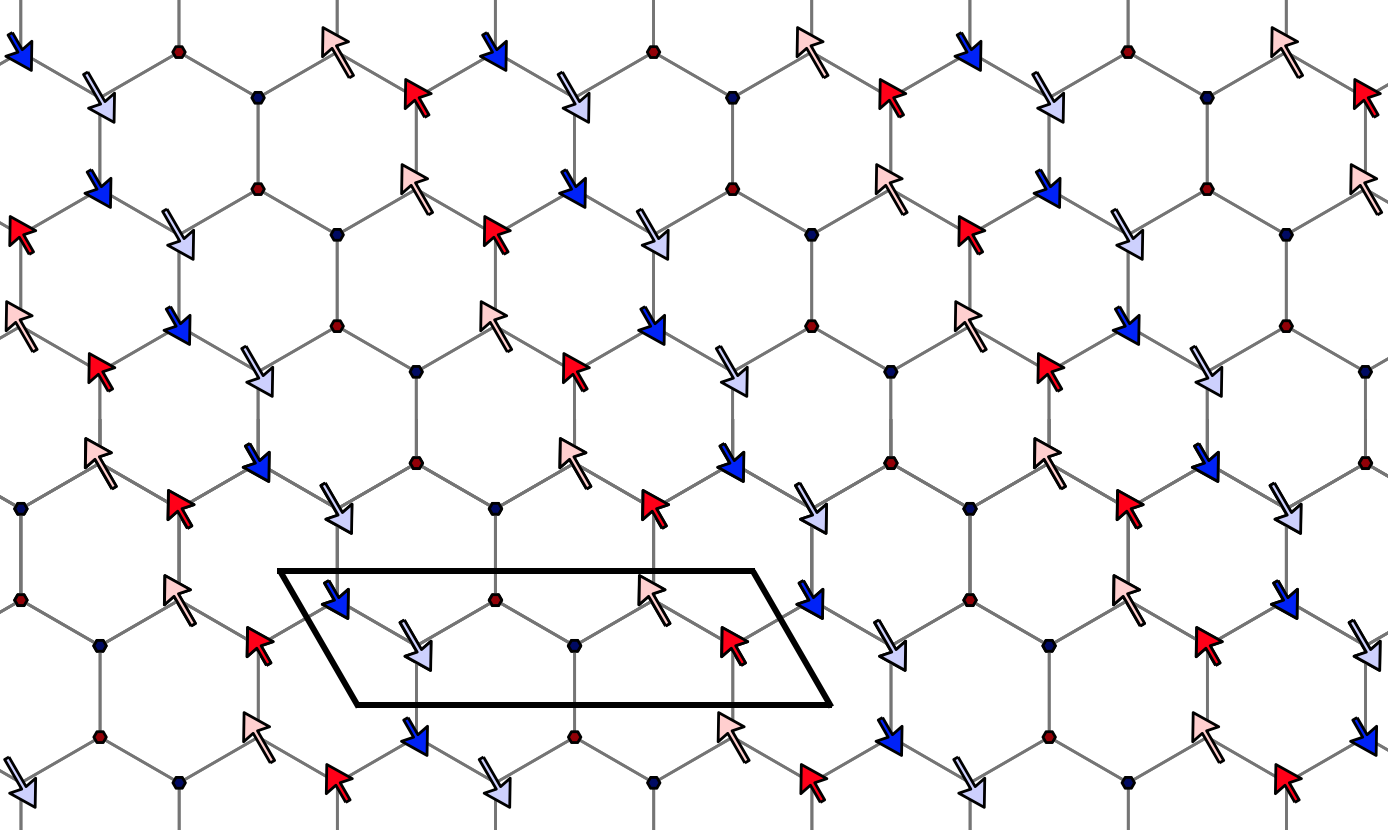}\caption{\label{fig:6classical} Pseudospin configuration at $\left(\theta/\pi,\,\phi/\pi,\,J_{B}/\bar{J}\right)=$
$\left(0.25,\,0.8,\,1/\sqrt{5}\right)$ located within the 6-site
phase.}
\end{figure}

\begin{figure}[H]
\centering{}\includegraphics[scale=0.18]{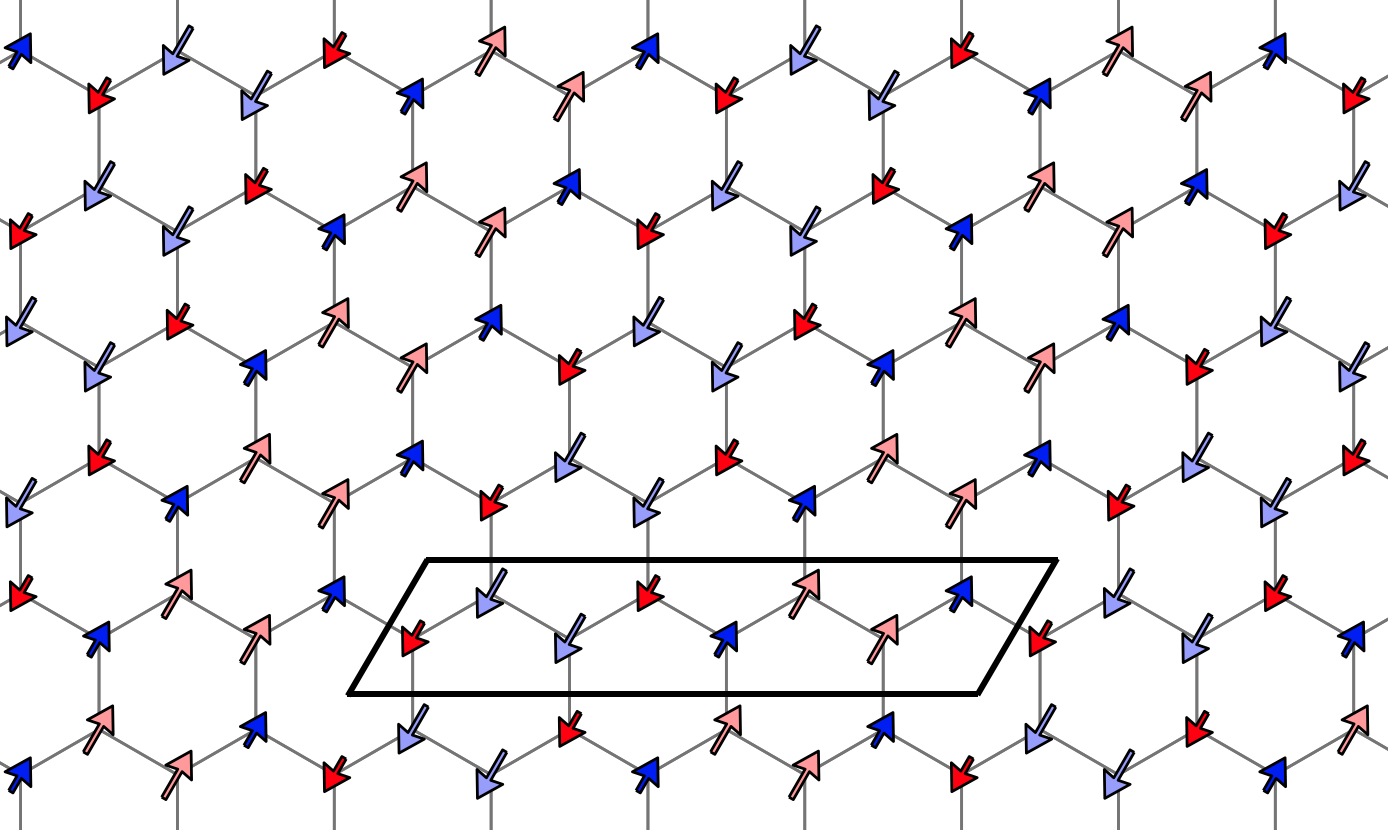}\caption{\label{fig:8Iclassical} Pseudospin configuration at $\left(\theta/\pi,\,\phi/\pi,\,J_{B}/\bar{J}\right)=$
$\left(0.375,\,0.175,\,1/\sqrt{5}\right)$ located within the $8_{I}$
phase.}
\end{figure}

\begin{figure}[H]
\centering{}\includegraphics[scale=0.18]{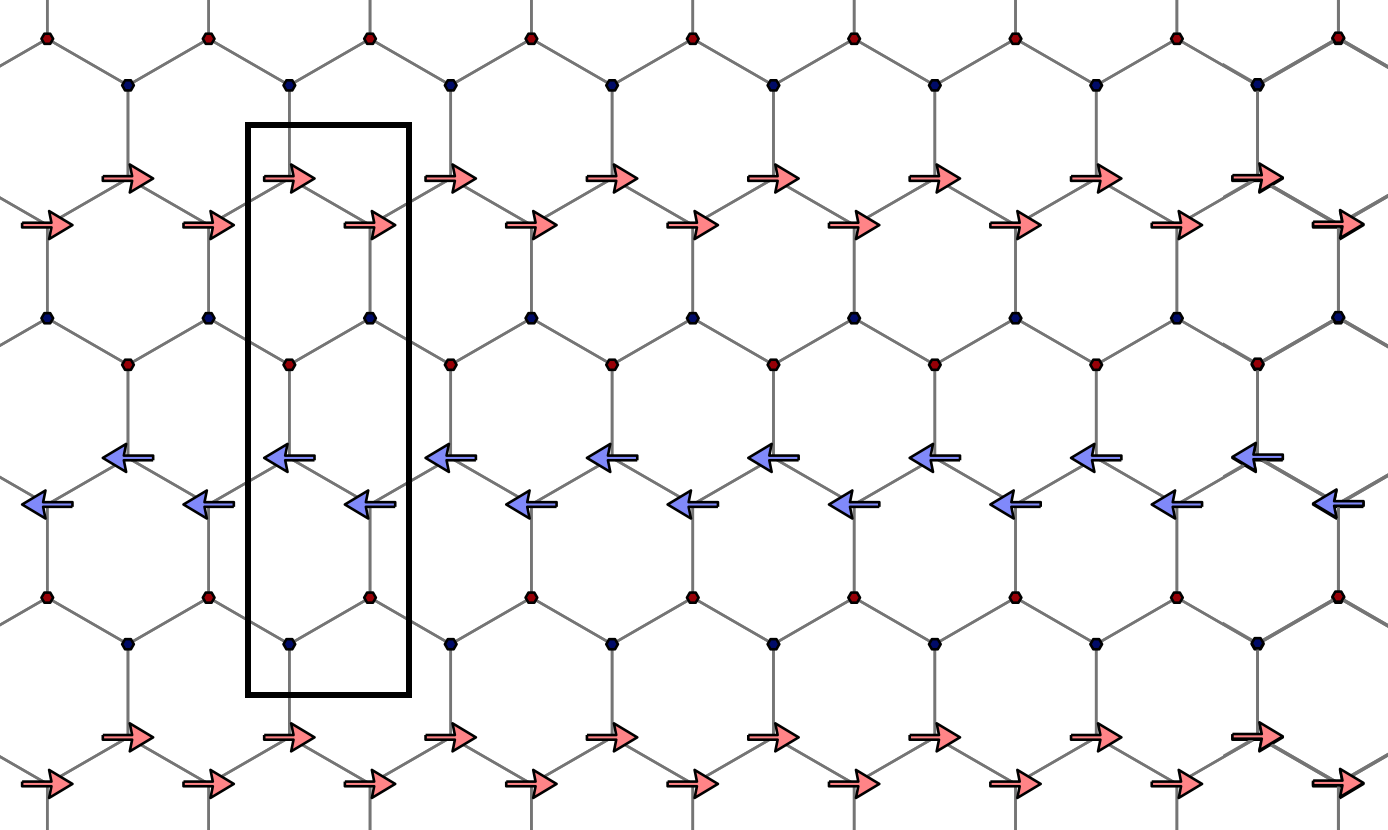}\caption{\label{fig:8IIclassical} Pseudospin configuration at $\left(\theta/\pi,\,\phi/\pi,\,J_{B}/\bar{J}\right)=$
$\left(0.35,\,0.765,\,1/\sqrt{5}\right)$ located within the $8_{II}$
phase.}
\end{figure}

\begin{figure}[H]
\centering{}\includegraphics[scale=0.18]{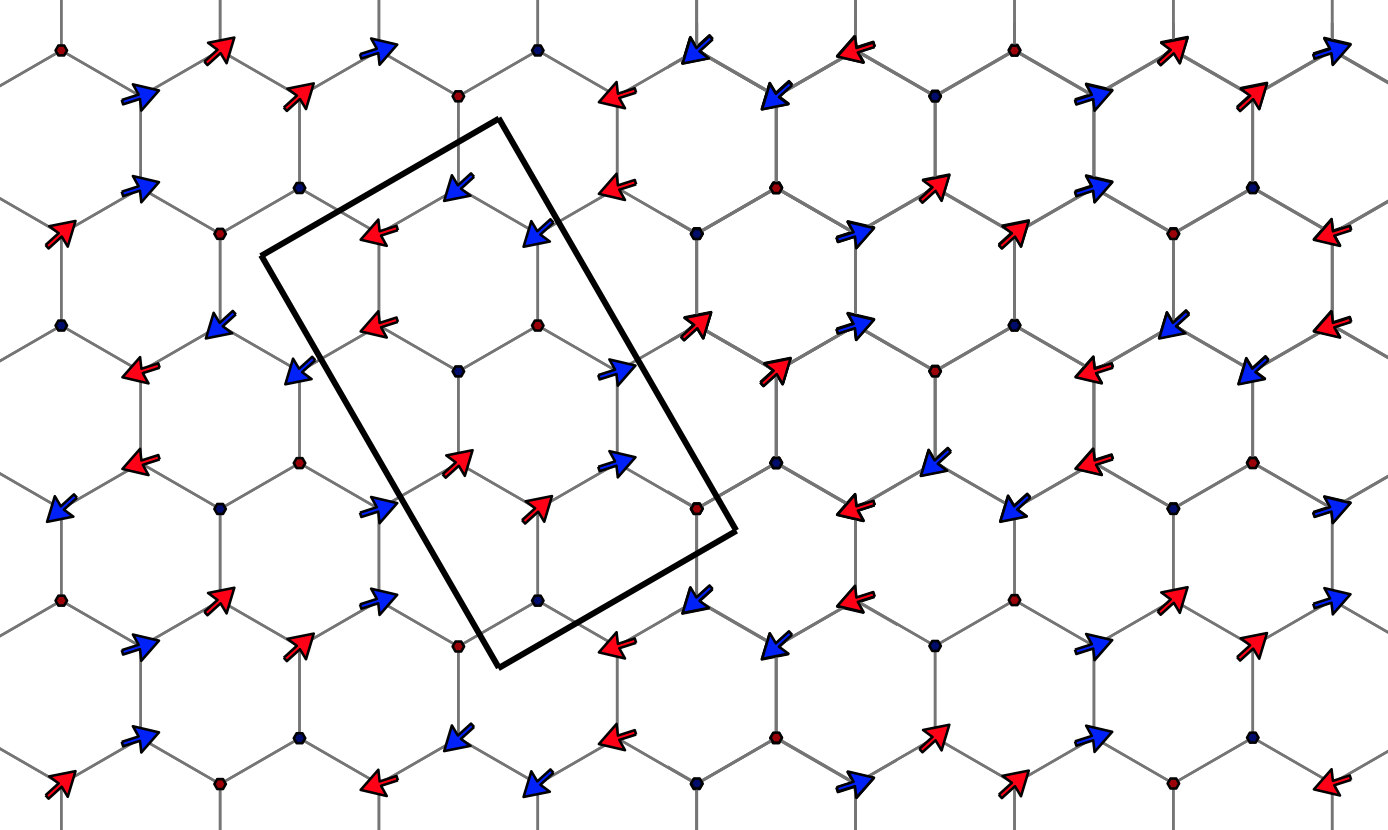}\caption{\label{fig:12Iclassical} Pseudospin configuration at $\left(\theta/\pi,\,\phi/\pi,\,J_{B}/\bar{J}\right)=$
$\left(0.25,\,0.72,\,1/\sqrt{5}\right)$ located within the $12_{I}$
phase.}
\end{figure}

\begin{figure}[H]
\centering{}\includegraphics[scale=0.18]{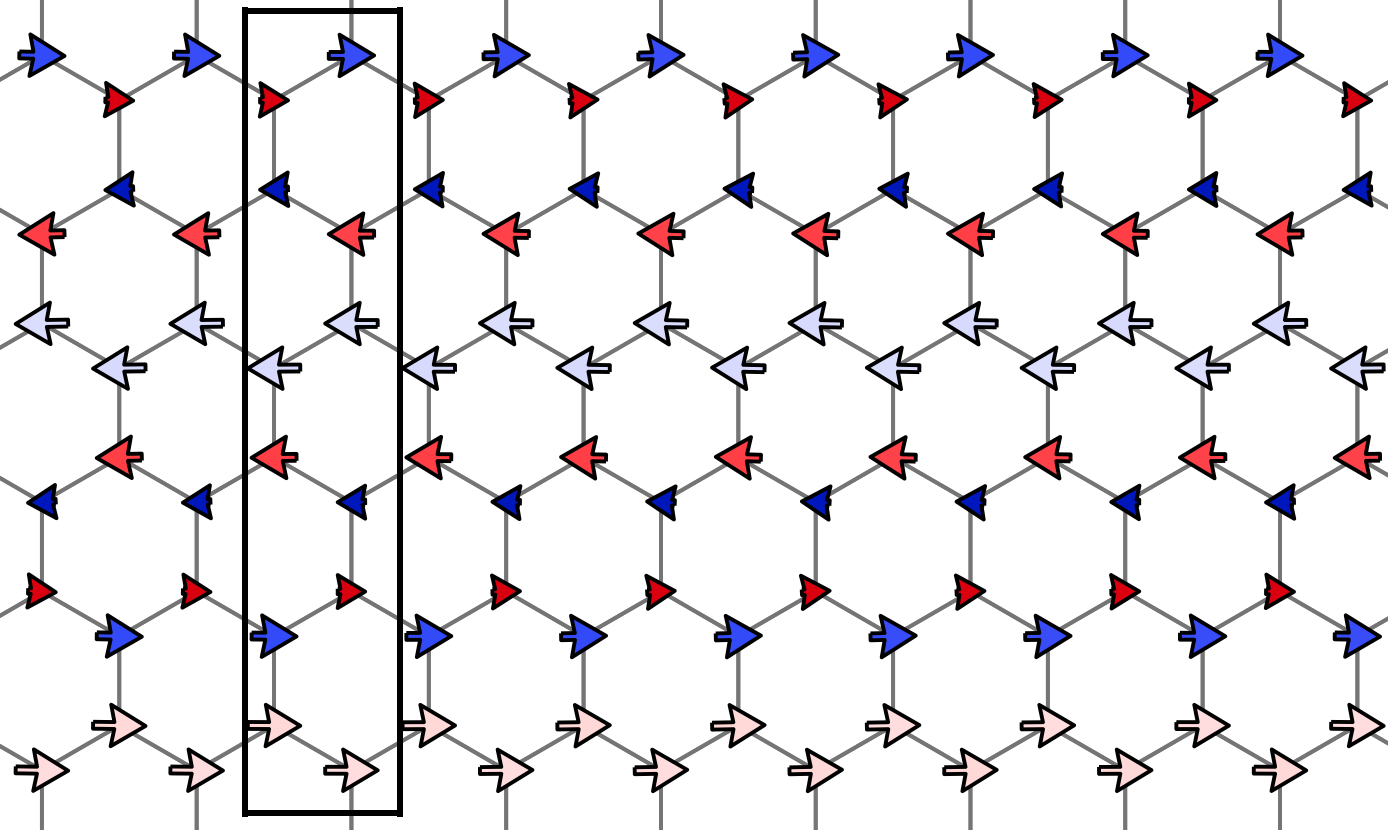}\caption{\label{fig:12IIclassical} Pseudospin configuration at $\left(\theta/\pi,\,\phi/\pi,\,J_{B}/\bar{J}\right)=$
$\left(0.5,\,0.81,\,1/\sqrt{5}\right)$ located within the $12_{II}$
phase.}
\end{figure}

\begin{figure}[H]
\centering{}\includegraphics[scale=0.18]{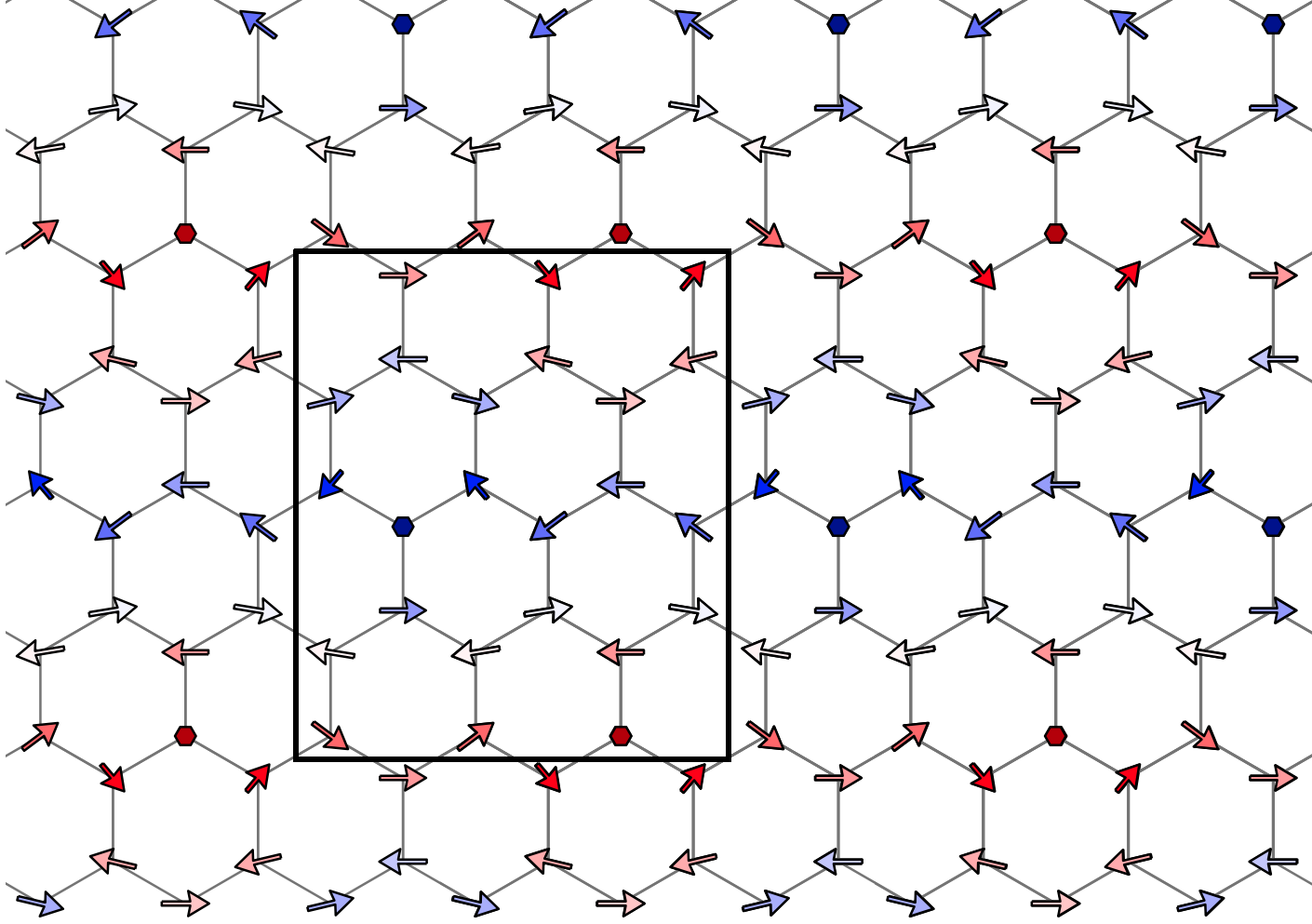}\caption{\label{fig:24LUCclassical} Pseudospin configuration at $\left(\theta/\pi,\,\phi/\pi,\,J_{B}/\bar{J}\right)=$
$\left(0.04,\,0.5,\,1/\sqrt{5}\right)$ located within the 24-site
phase.}
\end{figure}

\begin{figure}[H]
\centering{}\includegraphics[scale=0.18]{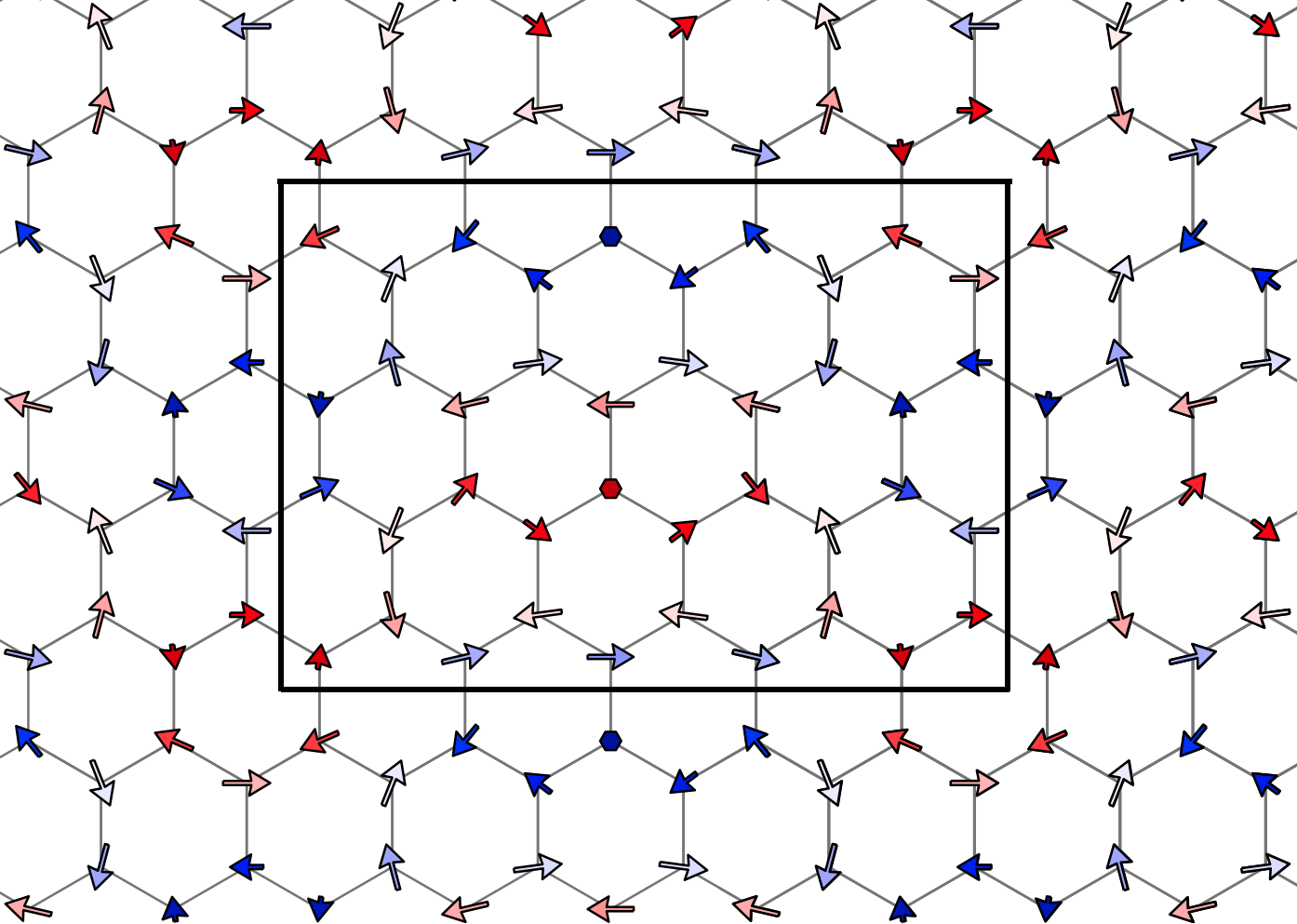}\caption{\label{fig:40LUCclassical} Pseudospin configuration at $\left(\theta/\pi,\,\phi/\pi,\,J_{B}/\bar{J}\right)=$
$\left(0.0,\,0.0,\,1/\sqrt{5}\right)$ located within the 40-site
phase.}
\end{figure}

Let us recall that the classical phase diagrams in Fig. 3 of the main
text are parameterized by $\left(\theta,\,\phi,\,J_{B}\right)$, where
$J_{\tau}=\bar{J}\,\text{cos }\theta,$ $J_{Q}=\bar{J}\,\text{sin }\theta\,\text{cos }\phi,$
and $J_{O}=\bar{J}\,\text{sin }\theta\,\text{sin }\phi$ at fixed
$h_{\text{eff}}=0$ and either $J_{B}=0\text{ or }\bar{J}/\sqrt{5}$
for Fig. 3(a) or 3(b), respectively. In the classical limit, each
pseudospin $\mathbf{s}_{i}=s_{i}^{a}\,\hat{\mathbf{e}}_{a}+s_{i}^{b}\,\hat{\mathbf{e}}_{b}+s_{i}^{c}\,\hat{\mathbf{e}}_{c}$
is then treated as an $O(3)$ vector satisfying $\left|\mathbf{s}_{i}\cdot\mathbf{s}_{i}\right|=1$;
the ground state can then be obtained by Monte Carlo simulated annealing
\citep{metropolis53,kirkpatrick1983optimization,kirkpatrick1984optimization},
see Appendix A of Ref. \citep{Rayyan_2021} for simulation details.
For a given pseudospin $\mathbf{s}_{i},$ Eq. (1) of the main text suggests that a
finite in-plane component $s_{i}^{a}\,\hat{\mathbf{e}}_{a}+s_{i}^{b}\,\hat{\mathbf{e}}_{b}$
corresponds to quadrupolar moments whereas a finite out-of-plane component
$s_{i}^{c}$ corresponds to the octupolar moment. In Figs. \ref{fig:afqclassical}-\ref{fig:40LUCclassical}
we present pseudospin configurations of each ordered phase that appears
in Fig. 3 of the main text along with the ordered unit cell. The size
of each arrow represents the in-plane component whereas the color
lying between red and blue represents the angle made with the out-of-plane
direction $\hat{\mathbf{e}}_{c}$, given by inverting $\text{cos }\theta_{\hat{\mathbf{e}}_{c}}=\mathbf{s}\cdot\hat{\mathbf{e}}_{c}=s^{c}$.
The phases in Figs. \ref{fig:afqclassical}-\ref{fig:vqclassical}
feature either quadrupolar or octupolar moments but not both, whereas
the phases in Figs. \ref{fig:zzclassical}-\ref{fig:40LUCclassical}
each contain both quadrupolar and octupolar ordering which are stabilized
due to the finite quadrupole-octupole interaction $J_{B}$.

\subsection{Phases in the quantum model}

\begin{figure}[H]
\includegraphics[scale=0.24]{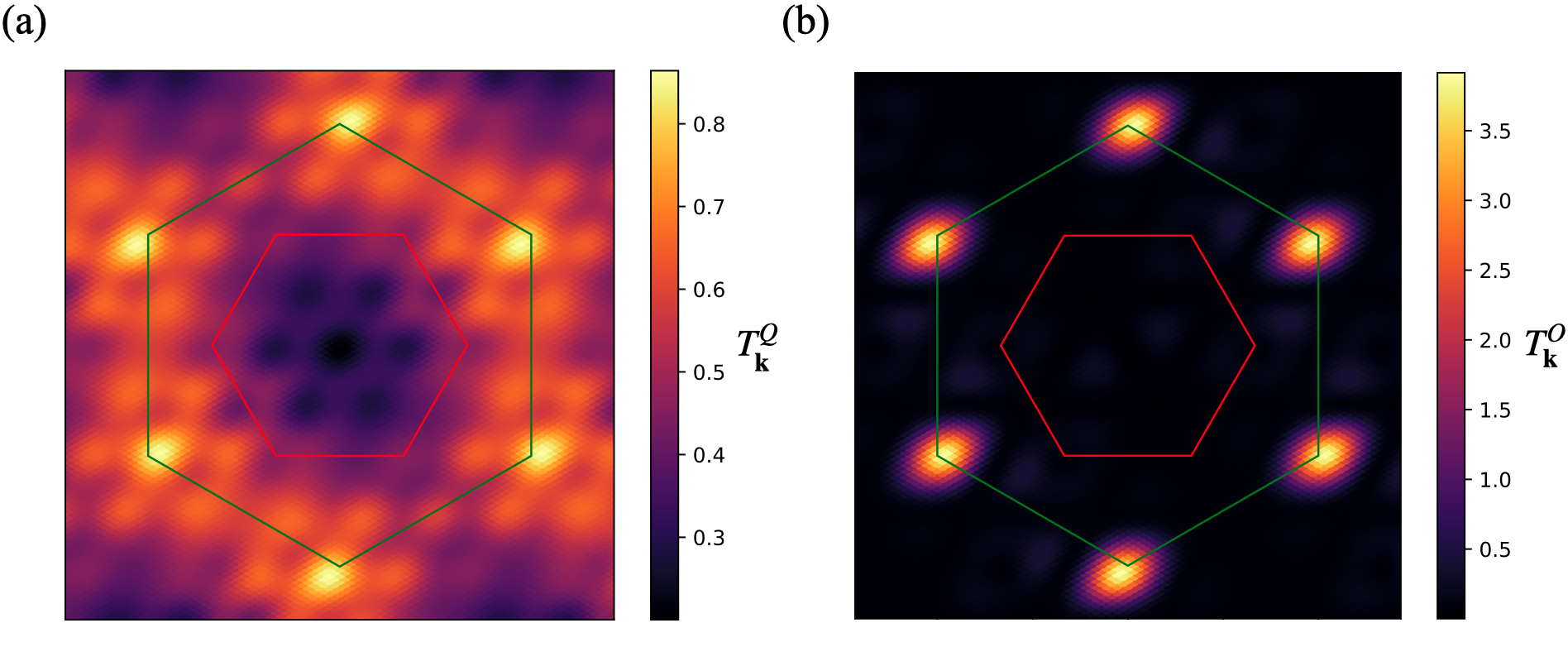}\caption{\label{fig:afoED} (a) Quadrupolar and (b) octupolar structure factors
of ground state obtained at $\left(\xi,\,h_{\text{eff}}\right)=$
$(-1.0,\,0.0)$ located within the AF$O$ phase. For Figs. \ref{fig:afoED}-\ref{fig:zzkmlED}
the $1^{\text{st}}$ and $2$$^{\text{nd}}$ crystal Brillouin zones
are shown in red and green respectively.}
\end{figure}

\begin{figure}[H]
\includegraphics[scale=0.24]{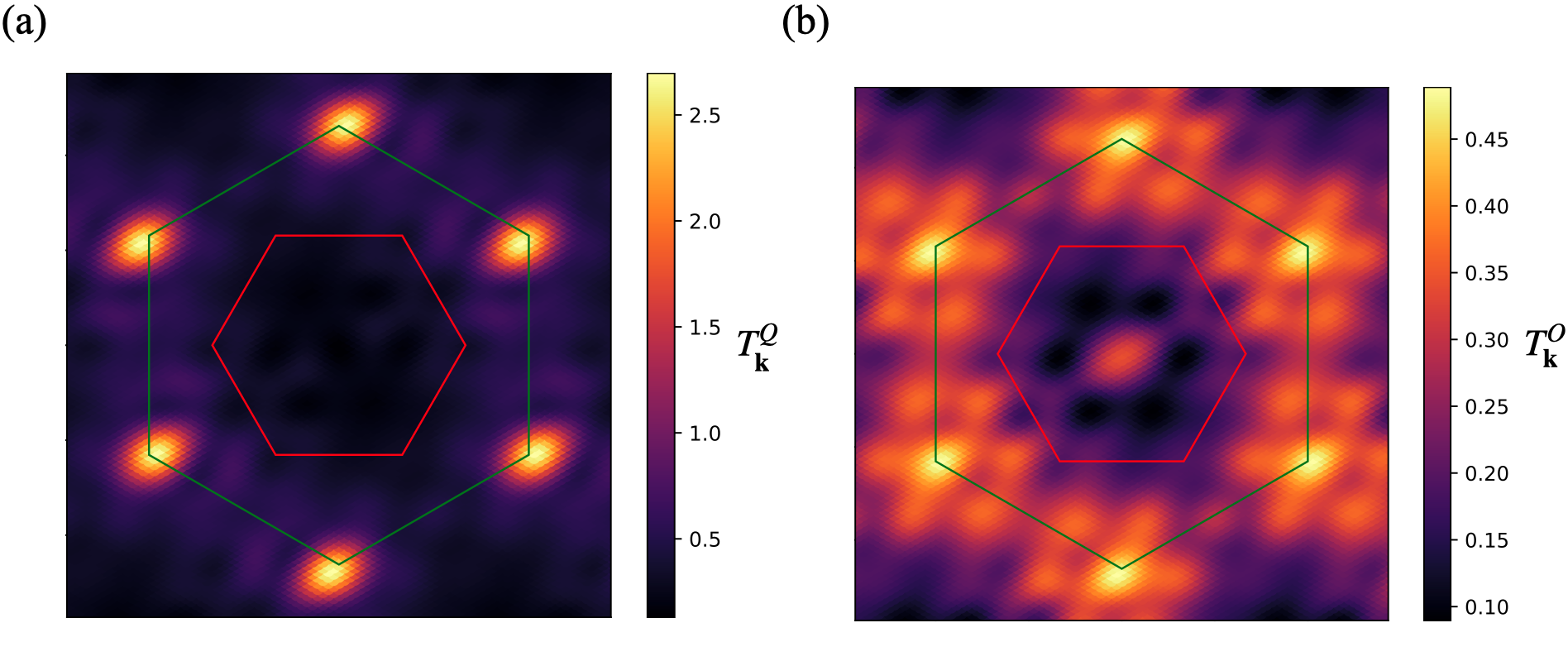}\caption{\label{fig:afqED}(a) Quadrupolar and (b) octupolar structure factors
of ground state obtained at $\left(\xi,\,h_{\text{eff}}\right)=$
$(-0.15,\,0.5)$ located within the AF$Q$ phase. The presence of
a small peak in $T_{\mathbf{k}}^{O}$ at the $\mathbf{k}=\Gamma$
point indicates the presence of ferro-octupolar correlations due to
the finite $h_{\text{eff}}$.}
\end{figure}

\begin{figure}[H]
\includegraphics[scale=0.24]{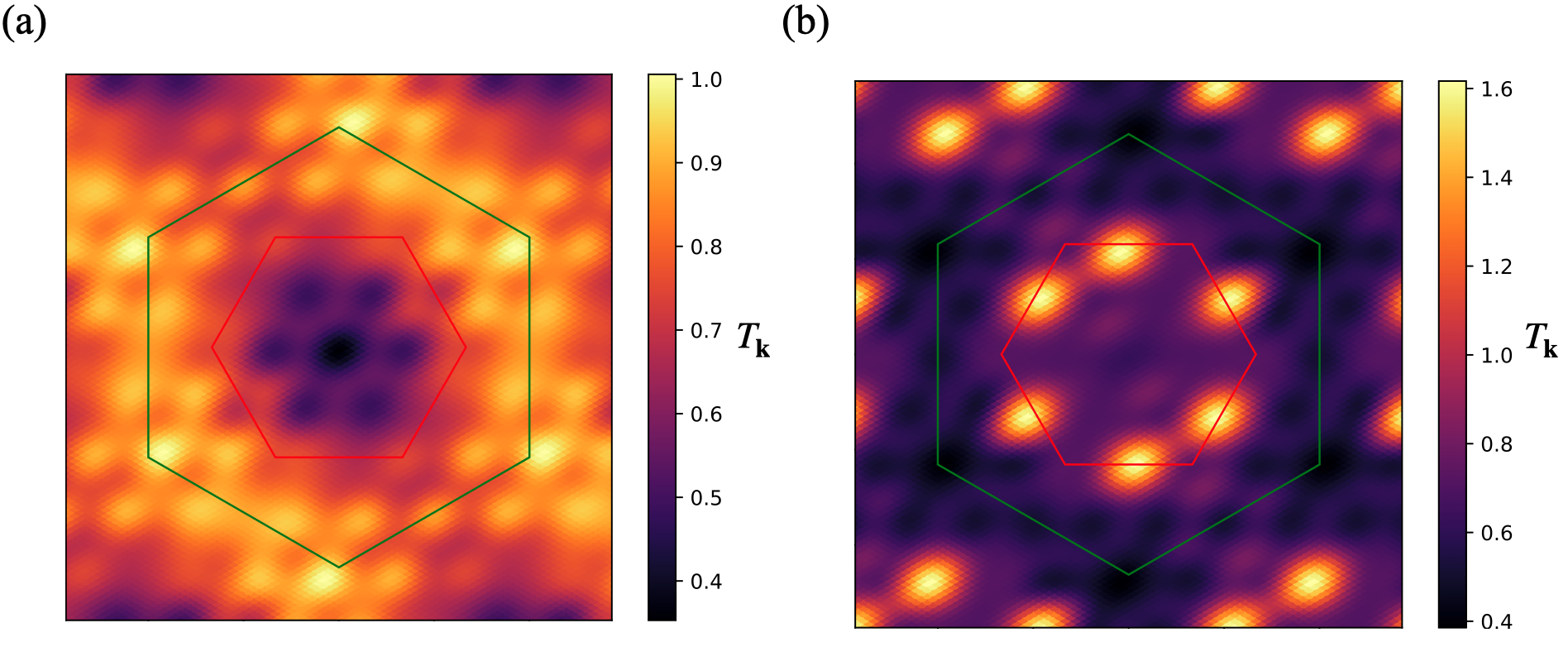}\caption{\label{fig:zzkmlED}Combined quadrupolar and octupolar structure factor
$T_{\mathbf{k}}=T_{\mathbf{k}}^{Q}+T_{\mathbf{k}}^{O}$ at (a) $\left(\xi,\,h_{\text{eff}}\right)=$
$(0.0,\,0.0)$ located within the KML phase, and (b) $\left(\xi,\,h_{\text{eff}}\right)=$
$(1.0,\,0.0)$ located within the ZZ phase.}
\end{figure}

Let us recall that the quantum phase diagram in Fig. 4 of the main
text is parameterized by $\left(\xi,h_{\text{eff}}\right)$ at fixed
$J_{Q}=0$ and $J_{\tau}=1$, where $\xi=\left(J_{B}-J_{O}\right)/\left(J_{B}+J_{O}\right)$
and is restricted to $-1\leq\xi\leq1$; in other words, tuning $\xi$
is equivalent to tuning the ratio $J_{B}/J_{O}=\left(1+\xi\right)/(1-\xi)$.
The $\mathcal{H}\Phi$ package provides the real-space correlation
functions $\braket{s_{i}^{\bar{\alpha}}s_{j}^{\bar{\beta}}}=\bra{\Psi}s_{i}^{\bar{\alpha}}s_{j}^{\bar{\beta}}\ket{\Psi}$
for $\bar{\alpha},\bar{\beta}\in\left\{ a,b,c\right\} $, where $\ket{\Psi}$
is the ground state of the pseudospin model and $i,j$ are sites on
the $24$-site cluster \citep{Kawamura_2017}. Let us define the quantity
\begin{equation}
T_{\mathbf{k}}^{\bar{\alpha}\bar{\beta}}=\frac{1}{N}\sum_{ij}\braket{s_{i}^{\bar{\alpha}}s_{j}^{\bar{\beta}}}e^{-i\mathbf{k}\cdot\left(\mathbf{r}_{i}-\mathbf{r}_{j}\right)}
\end{equation}
in analogy with the spin structure factor. The octupolar and quadrupolar
structure factors are then given by $T_{\mathbf{k}}^{O}\equiv T_{\mathbf{k}}^{cc}$
and $T_{\mathbf{k}}^{Q}\equiv T_{\mathbf{k}}^{aa}+T_{\mathbf{k}}^{bb}$,
and a signature of a long-range ordered phase are sharp peaks at some
$\mathbf{k}=\mathbf{k}^{*},$ where $\mathbf{k}^{*}$ are the corresponding
ordering wavevectors. In the quantum phase diagram we find fours phases:
AF$O$, AF$Q$, ZZ, and KML. The first three are ordered with ordering
wavevectors at $\mathbf{k}=\Gamma'$ for both AF$O$ and AF$Q$ and
$\mathbf{k}=M$ for the ZZ phase, see Figs. \ref{fig:afoED}-\ref{fig:zzkmlED}.
On the other hand, the KML is characterized by broad features in both
$T_{\mathbf{k}}^{Q}$ and $T_{\mathbf{k}}^{O}$, indicating a lack
of long-ranged multipolar order. Moreover, both AF$O$ and AF$Q$
feature sharp peaks in only one of $T_{\mathbf{k}}^{Q}$ and $T_{\mathbf{k}}^{O}$
as shown in Figs. \ref{fig:afoED} and \ref{fig:afqED}. On the other
hand, both the KML and ZZ phases feature a combination of quadrupolar
and octupolar correlations.
